\documentclass[journal]{IEEEtran}
\pdfoutput=1

\usepackage{cite}

\usepackage{amsmath}
\usepackage{algorithm}
\usepackage{algorithmic}
\usepackage{subfigure}
\usepackage{amssymb}
\usepackage{graphicx}
\usepackage{verbatim}
\usepackage{empheq}

\usepackage{array}
\usepackage{mathtools}
\usepackage{graphicx}
\usepackage{cases}
\usepackage[colorinlistoftodos]{todonotes}
\usepackage[colorlinks=true, linkcolor=black,citecolor=black,urlcolor=blue]{hyperref}
\usepackage{indentfirst}
\usepackage{multirow}
\usepackage{booktabs}
\usepackage{makecell}

\allowdisplaybreaks

\begin{document}

\title{Trajectory Design in Rechargeable UAV Aided Wireless Networks: A Unified Framework for Energy Efficiency and Flight Time Minimization}
\author{
Yuan~Liao and Vasilis~Friderikos, \IEEEmembership{Member,~IEEE} 

\thanks{The authors are with the Department of Engineering, King's College London, London WC2R 2LS, U.K (e-mail: yuan.liao@kcl.ac.uk; vasilis.friderikos@kcl.ac.uk).}
}


\maketitle

\begin{abstract}
This paper studies a rechargeable unmanned aerial vehicle (UAV) assisted wireless network, where a UAV is dispatched to disseminate information to a group of ground terminals (GTs) and returns to a recharging station (RS) before the on-board battery is depleted. The central aim is to design a UAV trajectory with the minimum total time duration, including the flight and recharging time, by optimizing the flying velocity, transmit power and hovering positions jointly. A flow-based mathematical programming formulation is proposed to provide optimal for joint optimization of flying and recharging time. Furthermore, to attack the curse of dimensionality for optimal decision making, a two-step method is proposed. In the first step, the UAV hovering positions is fixed and initialize a feasible trajectory design by solving a travelling salesman problem with energy constraints (TSPE) problem. In the second step, for the given initial trajectory, the time consumption for each sub-tour is minimized by optimizing the flying velocity, transmit power and hovering positions jointly. Numerical results show that the proposed method outperforms state of art techniques and reduces the aggregate time duration in an efficient way.
\end{abstract}

\begin{IEEEkeywords}
Unmanned aerial vehicle (UAV), 5G, wireless networks, trajectory design, energy efficiency, flight time minimization, integer programming, convex optimization.
\end{IEEEkeywords}

\IEEEpeerreviewmaketitle

\section{Introduction}
\label{introduction}

\IEEEPARstart{U}nmanned aerial vehicles (UAVs) are expected to play a significant role in next generation cellular networks due to their inherent high flexibility and controllable mobility. Various application scenarios of UAV-enable wireless communications, including aerial base stations, data collection/dissemination, mobile edge computing (MEC) and secrecy communication, have attracted intensive attention recently \cite{zeng2019accessing}, \cite{cao2018airborne}. However, because of the limited capacity of the on-board battery and the delay-tolerance of the data transmission, the aspects of energy efficiency and mission time minimization are two key challenges in UAV-enabled networks. This paper mainly captures a practical coupling of these two challenges and  proposes a unified framework to deal with the trade-off between them.

To start with, we first present in the sequel an intuitive illustration as a motivation statement to depict the coupling between energy efficiency and flight time minimization. Let us consider a wireless network where a rechargeable UAV disseminates information via wireless packet transmission to a group of ground terminals (GTs). Intuitively, when the UAV flies at the maximum attainable speed to serve each GT, it would achieve the shortest round-trip flight time but consume more energy. Thus, the UAV would require more time in replenishing when returning back at the depot. Conversely, an energy-efficient speed would result in a shorter recharging time but a longer flying duration. To deal with the trade-off between these two objectives, we aim to minimize the total time consumption, defined as the sum of flight and recharging time, by developing a novel trajectory planning in this paper.

Several existing research works concentrated on the energy-efficient trajectory design since the on-board battery limitation is, undoubtedly, one of the key challenges in UAV-assisted wireless networks. In \cite{mozaffari2016mobile}, a swarm of UAVs is applied to collect information from a group of moving ground devices, in which system the total energy consumption is optimized with the satisfaction of serving requirements. The time deadline is considered jointly with the data collection in \cite{9322626}. Moreover, the collection task completed by a single UAV is investigated in \cite{zhan2017energy}, \cite{yang2018energy}. In \cite{alzenad20173}, the transmit power of a UAV base station is minimized by adjusting its 3-D location. The ratio of the transmission rate and the energy consumption is optimized in \cite{zeng2017energy} by trajectory planning. A similar problem but with a rotary-wing UAV is studied in \cite{zeng2019energy}. The energy-aware trajectory is designed by a deep reinforcement learning approach in \cite{liu2018energy}. The energy-efficient connection of the backhual channels is discussed in \cite{monwar2018optimized}. Furthermore, the energy-aware network with non-orthogonal multiple access (NOMA) transmission is studied in \cite{zhang2020resource}, \cite{xi2019energy}. In \cite{manzoor2021ruin}, the energy-efficient resource allocation in a UAV-assisted cellular network is studied under the fifth generation (5G) new radio (NR) access technology.

The high mobility has always seen as the most attractive advantage of the UAV-assisted wireless network. To exploit this insight in a more detailed manner, several existing papers focus on the flight time minimization. In \cite{mozaffari2017wireless}, the hovering time of UAVs is minimized with the guarantee of the load requirements being satisfied. In \cite{gong2018flight}, the total mission time, including flight and hovering time, is optimized in a special application scenario where the ground devices are located across a linear segment. In \cite{zeng2018trajectory}, the UAV is dispatched to disseminate a common file to a group of terminals, and the problem has been formulated inline with the travelling salesman problem (TSP). The mission time minimization for the UAV-assisted data collection task is studied in \cite{li2019joint}. In \cite{yao2019qos2}, the UAV returns to a a depot when the on-board battery runs out, and the underlying trajectory is optimized to minimize the flight time. Moreover, the energy and flight time issues are studied jointly as a multi-objective optimization problem in \cite{zhan2020completion}, in which the Pareto optimality frontier is also discussed.

Unlike the multi-objective optimization problem proposed in \cite{zhan2020completion}, we propose hereafter an alternative view on the problem where we aim to unify these two core evaluations, i.e. the energy efficiency and the flight time minimization. Specifically, in this paper, a rechargeable UAV is dispatched to disseminate information to a group of GTs, so that the aggregate time consumption, i.e., the flight time as well as the recharging time, is minimized by proposing a novel trajectory planning. Although the wireless network assisted by a rechargeable UAV is investigated in \cite{yao2019qos2}, the recharging duration is ignored in that previous research work. Note that the consideration of recharging time is the key point to unify the energy and flight time issues since the positive correlation between the energy consumption and recharging time. To this end, the main contributions of this paper are summarized as follows,

\begin{itemize}
  \item [1)] 
  Unlike the previous work \cite{zeng2018trajectory}, in which the nominal TSP is applied to a UAV-enable network, the system assisted by a rechargeable UAV posses several further difficulties  to formulate due to the following two main reasons. First, in nominal TSP, each vertex is limited to be visited only once. However, in our problem, all GTs are to be served once but the recharging station (RS) is allowed to be visited several times for replenishment. Secondly, the on-board battery could not run out during the flying period. To this end, a novel flow-based constraint set is formulated to satisfy these two requirements.
  \item [2)]
  The formulated problem is notoriously difficult to attack due to its combinatorial nature (integer variables)  and non-convexity. To tackle this challenging problem, we propose a two-step method. In the first step, the UAV hovering points are fixed at the GT horizontal locations, and we solve a travelling salesman problem with energy constraints (TSPE)  to obtain an initial feasible trajectory design. In the second step, based on the initial trajectory, we minimize the time consumption for each sub-tour by optimizing the flying velocity, transmit power and hovering positions jointly.
  \item [3)]
  Extensive numerical investigations reveal that the two objectives, i.e. energy efficiency and flight time minimization, are coupled with each other. In other words, optimizing any of them alone cannot achieve the minimal total time consumption when the recharging process is taken into account. Besides, we also show that the proposed flow-based formulation for the TSPE problem has a significant gain compared to a recent closely relevant formulation proposed in \cite{yao2019qos2}.
\end{itemize}

The remainder of this paper is organized as follows. The UAV-enable wireless network model and the problem formulation are introduced in section \ref{systemmodel}. The proposed two-step method, together with the initialization and optimization procedures, are investigated in section \ref{initializing} and section \ref{VTPandHPopt}, respectively. Section \ref{NumericalResults} presents the simulation results and finally, section \ref{conclusion} concludes this paper.

\vspace{-0.2cm}
\section{System Model and problem formulation}
\label{systemmodel}

Hereafter, we consider a wireless communication system with $K$ ground terminals (GT) denoted by the set $\mathcal{K} = \{1,2,...,K\}$, the horizontal coordinates of which are known and fixed at $\mathbf{w}_i \in \mathbb{R}^2, \, \forall i \in \mathcal{K}$. A rotary-wing UAV acts as a carrier to transmit pre-cached data packets to GTs \cite{xu2018overcoming}. The UAV is powered by an on-board battery with limited capacity, and the battery can be replenished at the RS with a fixed location $\mathbf{w}_0$.

\vspace{-0.3cm}
\subsection{Hovering Model}
\label{HoveringModel}

In this paper, the fly-hover-communication protocol proposed in \cite{zeng2019energy} is adopted for the communication between UAV and GTs, in which the UAV successively visits $K$ hovering locations denoted by $\mathbf{h}_i \! \in \! \mathbb{R}^2, \forall i  \in  \mathcal{K}$, and transmits data to GTs only when it is hovering at location $\mathbf{h}_i$. Without loss of generality, the UAV is assumed to fly at a fixed altitude $H$ during the flight. Accordingly, the achievable rate $R_i$ for the wireless communication between UAV and the GT $i$ is assumed to follow the free-space path loss model and can be calculated as,
\begin{equation}
\label{AchRate}
\begin{aligned}
R_i = B \log_2 \Big( 1 + \frac{ P^{t}_i \rho_0}{\sigma^2(H^2 + \lVert \mathbf{h}_i - \mathbf{w}_i \rVert^2)}\Big), 
\end{aligned} 
\end{equation}
where $B$ is the available bandwidth for communication, $ P^{t}_i$ is the transmission power for the GT $i$, $\rho_0$ is the reference channel power gain at $1m$ distance, $\sigma^2$ is the power of channel noise and $\lVert\cdot\rVert$ denotes the 2-norm for a vector. The fly-hover-communication protocol requires that the UAV would keep communicating with GTs until the downloading process is finished \cite{zeng2019energy}. Thus, the hovering duration when the UAV transmits data to GT $i$ can be calculated as,
\begin{equation}
\label{Hoveringtime}
\begin{aligned}
t^d_i-t^a_i = \frac{D_i}{R_i},
\end{aligned} 
\end{equation}
where $t^a_i$ and $t^d_i$ is the time when the UAV arrives at and departs from $\mathbf{h}_i$, respectively. $D_i$ is the size of the data to be disseminated  for GT $i$. Therefore, the energy consumption of the UAV when hovering at $\mathbf{h}_i$ and offloading data to GT $i$ can be calculated as,
\begin{equation}
\label{Hoveringenergy}
\begin{aligned}
E^{h}_i & = (t^d_i-t^a_i) \big( \eta P^{t}_i  \! +\!  P^{h}  \big) \\ 
& = \frac{D_i \big( \eta   P^{t}_i  \! +\!  P^{h} \big)}{B \log_2 \big( 1 + \frac{   P^{t}_i \rho_0}{\sigma^2(H^2 + \lVert \mathbf{h}_i - \mathbf{w}_i \rVert^2)}\big)} ,
\end{aligned} 
\end{equation}
where $ P^{h}$ is the hovering power of the UAV when keeping stationary in the air and $\eta$ is the coefficient of transmit power \cite{yao2019qos}.

Moreover, since we consider both flying duration and recharging time, we define a novel concept as virtual time consumption (VTC) for a certain hovering period, which includes not only the hovering duration but also the recharging time, that is,
\begin{equation}
\label{VirtualHover}
\begin{aligned}
T^{h}_i & = \underbrace{t^d_i-t^a_i}_{\text{hovering time}}  + \underbrace{\frac{E^{h}_i}{ P^{r}}}_{\text{recharging time}} \\
& = \frac{D_i\big(\eta   P^{t}_i \!+ \!  P^{h} + P^{r}\big)}{ P^{r} B \log_2 \bigg( 1 + \frac{   P^{t}_i \rho_0}{\sigma^2(H^2 + \lVert \mathbf{h}_i - \mathbf{w}_i \rVert^2)}\bigg)},
\end{aligned} 
\end{equation}
where $ P^{r}$ is the recharging power.

For notational convenience, we combine the GTs and RS as an extended set $\mathcal{K}^{a} = \{0\} \bigcup \mathcal{K}$, in which the RS is indexed by $0$. Naturally, we have the following three equations when the UAV visits the RS node,
\begin{equation}
\label{recharging}
\begin{aligned}
\mathbf{h}_0 = \mathbf{w}_0, \; E^{h}_0 = 0, \; T^{h}_0 = 0
\end{aligned} 
\end{equation}
Note that the charging duration $T^{h}_0$ is seen as $0$ since this factor has already been calculated in the VTC.

\vspace{-0.3cm}
\subsection{Flying Model}
\label{FlyingModel}

In this subsection we focus on the UAV flying period. In \cite{zeng2019energy}, the flying power of rotatory-wing UAV is calculated as,
\begin{equation}
\label{flyingpower}
\begin{aligned}
P^{f} = P_0\Big(1+\frac{3\lVert \mathbf{v}(t)\rVert^2}{U_{tip}^2} \Big) + \frac{P_iv_0}{\lVert \mathbf{v}(t)\rVert} + \frac{1}{2}d_0\rho sA\lVert \mathbf{v}(t)\rVert^3
\end{aligned} 
\end{equation}
where $\mathbf{v}(t) \in \mathbb{R}^2 $ denotes velocity of the UAV at time $t$. $P_0$ and $P_i$ represent blade profile power and induced power in hovering status, respectively. $U_{tip}$ is the tip speed of the rotor blade. $v_0$ denotes the mean rotor induced velocity when hovering. $d_0$ and $s$ are the fuselage drag ratio and rotor solidity, respectively. $\rho$ and $A$ denote the air density and rotor disc area, respectively. 

Suppose the UAV flies from $\mathbf{h}_i$ to $\mathbf{h}_j$ without any stop, and the timestamps when departing from $\mathbf{h}_i$ and arriving $\mathbf{h}_j$  are denoted by $t^d_i$ and $t^a_j$, respectively. Then, the energy consumption when the UAV moves from from $\mathbf{h}_i$ to $\mathbf{h}_j$ would be,
\begin{small}
\begin{equation}
\label{flyingenergy}
\begin{aligned}
E^{f}_{ij} & = \int_{t^d_i}^{t^a_j}  \! \!  P^{f} \mathrm{d}t \\
& = \int_{t^d_i}^{t^a_j}  \! \! \Big( P_0\Big(1+\frac{3\lVert \mathbf{v}(t)\rVert^2}{U_{tip}^2} \Big) + \frac{P_iv_0}{\lVert \mathbf{v}(t)\rVert}  + \frac{1}{2}d_0\rho sA\lVert \mathbf{v}(t)\rVert^3 \Big)\mathrm{d}t 
\end{aligned} 
\end{equation}
\end{small}Similar as the definition of VTC for the hovering duration, we can also define VTC for the flying period including both flying and recharging time, that is,
\begin{small}
\begin{equation}
\label{Virtualfly}
\begin{aligned}
T^{f}_{ij} & = \underbrace{t^a_j - t^d_i}_{\text{flying time}} + \underbrace{\frac{E^{f}_{ij}}{ P^{r}}}_{\text{recharging time}} \\
& = t^a_j - t^d_i +  \frac{1}{ P^{r}} \int_{t^d_i}^{t^a_j}  \! \! \Big( P_0\big(1+\frac{3\lVert \mathbf{v}(t)\rVert^2}{U_{tip}^2} \big) + \frac{P_iv_0}{\lVert \mathbf{v}(t)\rVert}  \\ 
& \quad  + \frac{1}{2}d_0\rho sA\lVert \mathbf{v}(t)\rVert^3 \Big)\mathrm{d}t 
\end{aligned} 
\end{equation}
\end{small}To eliminate the variables $t^a_j$ and $t^d_i$, we propose the following proposition,

\textit{\textbf{Proposition 1:} Without loss of optimality for minimizing $T^{f}_{ij}$, the UAV flies along the line segment between $\mathbf{h}_i$ and $\mathbf{h}_j$.}

\textit{Proof:} We prove this by induction. For any non-straight trajectory, denoted as $\mathcal{Q}_{ij}$, we can always find a corresponding straight trajectory $\mathcal{Q'}_{ij}$, the velocity in which has the similar value (2-norm) as the velocity in $\mathcal{Q}_{ij}$ at each time slot. Accordingly, it can be easily verified that $\mathcal{Q'}_{ij}$ must arrive $\mathbf{h}_j$ earlier than $\mathcal{Q}_{ij}$ and consumes less energy. Thus, the non-straight trajectory $\mathcal{Q}_{ij}$ always consumes more time and energy than the corresponding straight trajectory $\mathcal{Q'}_{ij}$. This completes the proof of Proposition 1. $\square$

Besides, to avoid an infinite number of variables, we make a necessary assumption as,

\textit{\textbf{Assumption 1:} The UAV keeps a constant speed when flying from $\mathbf{h}_i$ to $\mathbf{h}_j$, which is denoted as $V_{ij}$.}

Based on Proposition 1 and Assumption 1, the equation \eqref{flyingenergy} can be rewritten as,
\begin{small}
\begin{equation}
\label{flyingenergy2}
\begin{aligned}
E^{f}_{ij} & =  \frac{\lVert\mathbf{h}_i-\mathbf{h}_j\rVert}{V_{ij}} P^{f} \\
& = \frac{\lVert\mathbf{h}_i-\mathbf{h}_j\rVert}{V_{ij}} \Big( P_0\Big(1+\frac{3V_{ij}^{2}}{U_{tip}^2} \Big) + \frac{P_iv_0}{V_{ij}} + \frac{1}{2}d_0\rho sAV_{ij}^{3} \Big) \\
& = \lVert\mathbf{h}_i-\mathbf{h}_j\rVert \Big( \frac{1}{2}d_0\rho sAV_{ij}^{2} + \frac{3P_0V_{ij}}{U_{tip}^2} + \frac{P_0}{V_{ij}} + \frac{P_iv_0}{V_{ij}^2}  \Big) \\
& \triangleq \lVert\mathbf{h}_i-\mathbf{h}_j\rVert \Big( \psi_1 V_{ij}^{2} + \psi_2 V_{ij} + \frac{\psi_3}{V_{ij}} + \frac{\psi_4}{V_{ij}^2} \Big)
\end{aligned} 
\end{equation}
\end{small}where $\psi_1$, $\psi_2$, $\psi_3$ and $\psi_4$ are constants defined for notational convenience. Similarly, the equation \eqref{Virtualfly} can be rewritten as,
\begin{small}
\begin{equation}
\label{Virtualfly2}
\begin{aligned}
T^{f}_{ij} & =  \underbrace{\frac{\lVert\mathbf{h}_i-\mathbf{h}_j\rVert}{V_{ij}}}_{\text{flying time}} +  \underbrace{\frac{E^{f}_{ij}}{ P^{r}}}_{\text{recharging time}} \\
& = \lVert\mathbf{h}_i-\mathbf{h}_j\rVert \Big( \frac{d_0\rho sA}{2P^{r}}V_{ij}^{2} + \frac{3P_0V_{ij}}{P^{r}U_{tip}^2} + \frac{P_0+P^{r}}{P^{r}V_{ij}}+\frac{P_iv_0}{P^{r}V_{ij}^2} \Big)\\
& \triangleq \lVert\mathbf{h}_i-\mathbf{h}_j\rVert \Big( \tau_1 V_{ij}^{2} + \tau_2 V_{ij} + \frac{\tau_3}{V_{ij}} + \frac{\tau_4}{V_{ij}^2}  \Big)
\end{aligned}
\end{equation}
\end{small}where $\tau_1$, $\tau_2$, $\tau_3$ and $\tau_4$ are constants defined for convenience.

\vspace{-0.3cm}
\subsection{Flow-based constraint set for the TSPE problem}
\label{TSPEM}

Based on the aforementioned hovering and flying model, we now turn our focus on the order of UAV visits. We name our problem as the TSPE problem due to the energy and recharging considerations. Compared with the nominal TSP, there are two distinct differences that need to be pointed out. Firstly, in TSP, each vertex is limited to be visited once. However, in TSPE, all hovering points $\mathbf{h}_i$ are limited to be achieved at least once whilst the RS position $\mathbf{h}_0$ is allowed to be visited  several times for UAV recharging. In other words, the UAV trajectory is allowed to contain more than one sub-tours\footnote{In this paper, a sub-tour is defined as a Hamiltonian cycle, in which the first vertex is equal to the last vertex (closed trail). \cite{gross2005graph}} and all sub-tours are required to be overlapped at the location of RS. Secondly, in TSPE, we also need to guarantee that the on-board battery would not be depleted during the whole serving period.

To satisfy these two requirements, we propose a flow-based formulation \cite{gavish1978travelling}, \cite{sundar2016formulations}, that is,
\begin{subequations}
\begin{empheq}[left={\empheqlbrace\,}]{align}
& \sum_{i \in \mathcal{K} } x_{0i} = \sum_{i \in \mathcal{K} } x_{i0}, \label{TSPE1}\\
&  \sum_{i \in \mathcal{K}^a } x_{ij} = 1 , \; \forall j \in \mathcal{K},  \label{TSPE2} \\
& \sum_{i \in \mathcal{K}^a } x_{ji} = 1 , \; \forall j \in \mathcal{K},  \label{TSPE3} \\
&  \sum_{j \in \mathcal{K}^a } \! z_{ij} - \! \sum_{j \in \mathcal{K}^a } \! z_{ji} = \sum_{j \in \mathcal{K}^a} \! \big( E^{h}_i + E^{f}_{ij} \big) x_{ij} ,\forall i \in \mathcal{K},  \label{TSPE4} \\
& z_{0i} = E^{f}_{0i}x_{0i}, \; \forall i \in \mathcal{K},  \label{TSPE5}\\
& 0\leq z_{ij} \leq F^{max}x_{ij}, \; \forall (i,j) \in \mathcal{E} \label{TSPE6}\\
& x_{ij} \in \{0,1\}, \; \forall (i,j) \in \mathcal{E},  \label{TSPE7}
\end{empheq}
\end{subequations}
where by $\mathcal{E}$ we denote a set that includes all unordered pairs of vertices in $\mathcal{K}^a$, that is, $\mathcal{E} \triangleq \{(i,j)\big|i,j \! \in \! \mathcal{K}^a, i\neq j \}$, $F^{max}$ is the on-board battery capacity, $x_{ij}$ are binary variables and $x_{ij}=1$ represents that the edge $(i,j)$ is travelled by the UAV, whether it is from $i$ to $j$ or from $j$ to $i$, and $z_{ij}$ are the flow variables associated to all edges. Observe that \eqref{TSPE1}-\eqref{TSPE3} impose the degree constraints for all GTs and RS. Also, \eqref{TSPE7} reflects the binary restriction for the variable $x_{ij}$. The following two Lemmas illustrate how the constraints \eqref{TSPE4}-\eqref{TSPE6} operate for the two aforementioned TSPE guarantees.

\textit{\textbf{Lemma 1:} The constraint \eqref{TSPE4} guarantees that if a vertex set $\mathcal{K}^{a,s} \subseteq \mathcal{K}^a $ constitutes a sub-tour, the RS must be included by $\mathcal{K}^{a,s}$, i.e. $0\in \mathcal{K}^{a,s}$.}

\textit{Proof:} We prove this by induction. Assume that there is a sub-tour without RS. Since the sub-tour is actually a cycle in the graph, we simply choose a GT $i_1  \in \mathcal{K}^{a,s}$ as the start and end point. Thus, all GTs in this $\mathcal{K}^{a,s}$ are visited in tandem and let us denoted by $i_1, i_2, ... , i_r, i_1$. Now let us define $z_{i_1i_2} = f$. According to the \eqref{TSPE4}, it follows that,  $z_{i_2i_3} = f + E^{h}_{i_2} + E^{f}_{i_2i_3}$, ..., $z_{i_ri_1} = f + \sum_{k=2}^{r-1} E^{h}_{i_k} + \sum_{k=2}^{r-1}E^{f}_{i_ki_k+1} + E^{h}_{i_r} + E^{f}_{i_ri_1}$. Therefore, we have $z_{i_1i_2} - z_{i_ri_1} = - (\sum_{k=2}^{r-1} E^{h}_{i_k} + \sum_{k=2}^{r-1}E^{f}_{i_ki_k+1} + E^{h}_{i_r} + E^{f}_{i_ri_1})$. Alternatively, we note that the above is actually equal to the following equation, 
\begin{small}
\begin{equation}
\label{contradict}
\begin{aligned}
\sum_{j \in \mathcal{K}^a } \! z_{i_1j} - \! \! \sum_{j \in \mathcal{K}^a } \! z_{ji_1}\! = \! -\big(\sum_{k=2}^{r-1} \! E^{h}_{i_k} + \! \sum_{k=2}^{r-1}E^{f}_{i_ki_k+1} + E^{h}_{i_r} + E^{f}_{i_ri_1} \big)
\end{aligned}
\end{equation}
\end{small}which contradicts with \eqref{TSPE4}. Thus, a sub-tour must involve the RS. This completes the proof of Lemma 1.
$\square$

\textit{\textbf{Lemma 2:} The constraints \eqref{TSPE4}-\eqref{TSPE6} guarantee that the on-board battery would not run out during the serving period.}

\textit{Proof:} According to Lemma 1, the RS must be involved in a sub-tour, i.e., $0\in \mathcal{K}^{a,s}$. Since the sub-tour is a Hamiltonian cycle, we can select the RS as the start and end point without loss of generality. In other words, the UAV would visit $0,i_1,i_2,...,i_r,0$ in turn. From \eqref{TSPE5}, we have $z_{0i_1} = E^{f}_{0i_1}$. From \eqref{TSPE4}, it follows $z_{i_1i_2} = E^{f}_{0i_1} + E^{h}_{i_1} + E^{f}_{i_1i_2}$, ... , $z_{i_r0} = E^{f}_{0i_1} + \sum_{k=2}^{r-1} E^{h}_{i_k} + \sum_{k=2}^{r-1}E^{f}_{i_ki_k+1} + E^{h}_{i_r} + E^{f}_{i_r0}$. In other words, the flow variable $z_{ij}$ can be seen as the energy consumption when the UAV departs from $j$, and the predecessor vertex of $j$ is $i$. The constraints in \eqref{TSPE6} guarantee that the energy consumption can be neither lower than $0$ nor greater than $F^{max}$. This completes the proof of Lemma 2. $\square$

\vspace{-0.3cm}
\subsection{Problem formulation}

According to the proposed system model and the aforementioned preliminaries, the trajectory design problem can be formulated as follows,
\begin{align}
\mathrm{(P1):} \quad &  \min_{\mathbf{H}, \mathbf{P}, \mathbf{V},\mathbf{X}, \mathbf{Z} } \;  \sum_{i \in \mathcal{K} } T^{h}_i +  \sum_{(i,j) \in \mathcal{E} } x_{ij} T^{f}_{ij} \label{Pro1}\\
s.t.
\quad & R_i \geq R^{th}, \; \forall i \in \mathcal{K} \tag{\ref{Pro1}{a}} \label{Pro1C1}\\
\quad & 0 \leq   P^{t}_i \leq P^{max},  \;  \forall i \in \mathcal{K}  \tag{\ref{Pro1}{b}} \label{Pro1C2}\\
\quad & 0 \leq V_{ij} \leq V^{max} \tag{\ref{Pro1}{c}} , \;  \forall (i,j) \in \mathcal{E} \label{Pro1C3}\\
\quad &  \eqref{recharging}, \; \eqref{TSPE1}-\eqref{TSPE7} \tag{\ref{Pro1}{d}} \label{Pro1C4}
\end{align}
where $\mathbf{H} \triangleq \{\mathbf{h}_i \big| \, \forall i\in \mathcal{K}^a \}$, $\mathbf{P} \triangleq \{  P^{t}_i \big | \, \forall i \in \mathcal{K} \}$, $\mathbf{V} \triangleq \{V_{ij} \big|\, \forall (i,j) \in \mathcal{E} \}$, $\mathbf{X} \triangleq \{x_{ij} \big | \, \forall (i,j) \in \mathcal{E} \}$ and $\mathbf{Z} \triangleq \{z_{ij} \big | \, \forall (i,j) \in \mathcal{E} \}$ are the set of variables. Constraints in \eqref{Pro1C1} relate to quality of service (QoS) support which guarantee that the downloading throughput must be greater than a pre-defined threshold $R^{th}$. The actual value of $R^{th}$ depends on different application scenarios \cite{zeng2019accessing}. Finally, constraints \eqref{Pro1C2} and \eqref{Pro1C3} limit the maximum transmit power and flying speed for the UAV, respectively.

The main difficulty for optimally solving (P1) lies in the following two reasons. Firstly, the trajectory variables $\mathbf{X}$ are binary so that \eqref{Pro1C4} involves integer constraints. Although several efficient tools can be used to solve the linear integer programming formulation for small to medium network instances, the nonlinear integer constraints in \eqref{TSPE4} cannot be handled directly. Secondly, even with given trajectory variables $\mathbf{X}$ and flow variables $\mathbf{Z}$, the objective function and the constraints \eqref{Pro1C1}, \eqref{TSPE4} and \eqref{TSPE5} are not convex with respect to the variables.

To handle this challenging optimization problem, we propose a two-step method as follows. In the first step, the hovering positions are fixed to the corresponding GT locations, i.e. $\mathbf{h}_i \!=\! \mathbf{w}_i$, which means the UAV would communicate with GTs when hovering right above them. We then optimize $\mathbf{V}$ and $\mathbf{P}$ separately. Based on the results, we solve the TSPE problem to obtain a feasible initial trajectory design. This step is named as initializing trajectory design and proposed in section \ref{initializing}. In the second step, with the given order of UAV visits, we optimize the $\mathbf{H}$, $\mathbf{P}$ and $\mathbf{V}$ jointly to minimize the total time. This step is presented in section \ref{VTPandHPopt}.

\vspace{-0.15cm}
\section{Initializing trajectory design}
\label{initializing}

In this section, to efficiently initialize a feasible trajectory, we start with the following assumption,

\textit{\textbf{Assumption 2:} In the step of trajectory initialization, the UAV would \\
\indent 1) hover right above the corresponding GT positions, that is, $ \{ \mathbf{h}_i \big| \, \mathbf{h}_i = \mathbf{w}_i, \, \forall i \in \mathcal{K}\}$. \\
\indent 2) keep a constant velocity, which is denoted as $V^{ini}$, during the flying period, that is,  $ \{ V_{ij} \big| \, V_{ij} = V^{ini}, \, \forall (i,j) \in \mathcal{E} \}$.}

Note that Assumption 2 is only held in this section for initializing a feasible trajectory. Introducing Assumption 2 into \eqref{AchRate}, \eqref{Hoveringenergy}-\eqref{VirtualHover} and  \eqref{flyingenergy2}-\eqref{Virtualfly2} and denoting the results as $R^{ini}_i$, $E^{h,ini}_i$, $T^{h,ini}_i$, $E^{f,ini}_{ij}$ and $T^{f,ini}_{ij}$, respectively, the problem (P1) can be rewritten as,
\begin{align}
\mathrm{(P2):} \; &  \min_{ V^{ini}, \mathbf{P},  \mathbf{X}, \mathbf{Z} } \;  \sum_{i \in \mathcal{K} } T^{h,ini}_i +  \sum_{(i,j) \in \mathcal{E} } x_{ij} T^{f,ini}_{ij} \label{Pro2}\\
s.t.
\quad & R^{ini}_i \geq R^{th}, \; \forall i \in \mathcal{K} \tag{\ref{Pro2}{a}} \label{Pro2C1}\\
\quad & 0 \leq V^{ini} \leq V^{max} \tag{\ref{Pro2}{b}} \label{Pro2C2}\\
\quad & \sum_{j \in \mathcal{K}^a } \! z_{ij} - \! \sum_{j \in \mathcal{K}^a } \! z_{ji}  = \sum_{j \in \mathcal{K}^a} \! \big( E^{h,ini}_i + E^{f,ini}_{ij} \big) x_{ij}, \forall i \in \mathcal{K}  \tag{\ref{Pro2}{c}} \label{Pro2C3}\\
\quad & z_{0i} = E^{f,ini}_{0i}x_{0i}, \; \forall i \in \mathcal{K} \tag{\ref{Pro2}{d}} \label{Pro2C4}\\
\quad &  \eqref{recharging}, \eqref{TSPE1}-\eqref{TSPE3}, \eqref{TSPE6}-\eqref{TSPE7}, \eqref{Pro1C2} \tag{\ref{Pro2}{e}} \label{Pro2C5}
\end{align}
Problem (P2) is still not solvable due to the integer variables in nonlinear objective function and constraints \eqref{Pro2C3}. To handle this problem, we decouple the variables to three groups, $V^{ini}$, $\mathbf{P}$ and $ (\mathbf{X}, \mathbf{Z})$, and solve them separately. It is worth pointing out that although this method can not achieve the optimal solution for (P2), it could provide a feasible trajectory efficiently for further optimization. Moreover, the simulation results in section \ref{NumericalResults} show that this decoupling strategy can at least achieve a locally optimal result.

\vspace{-0.3cm}
\subsection{Velocity Initialization}
\label{VelocityInitialization}
Since we assume that the UAV keeps a constant velocity $V^{ini}$ during the whole serving period in this section, the optimal value of $V^{ini}$ can be achieved by solving the following problem,
\begin{align}
\mathrm{(P3):} \quad & \min_{V^{ini}} \, \tau_1 V^{ini^{2}} + \tau_2 V^{ini} + \frac{\tau_3}{V^{ini}} + \frac{\tau_4}{V^{ini^2}} \label{Pro3}\\
s.t.
\quad &  0 \leq V^{ini}  \leq V^{max}  \tag{\ref{Pro3}{a}} \label{Pro3C1}
\end{align}
where the objective function of (P3) is the unit VTC for flying $1 \mathrm{m}$. Moreover, problem (P3) is convex since the objective function is convex with respect to the non-negative variable $V^{ini}$ and the constraint \eqref{Pro3C1} is linear inequality. Thus, the problem (P3) can be solved efficiently by standard convex optimization techniques. Denote the optimal solution of problem (P3) as $V^{ini^*}$.

\vspace{-0.3cm}
\subsection{Transmit Power Initialization}
\label{TransmitPowerInitialization}
When the UAV hovers right above and communicate with GT $i$, the problem for optimizing transmit power $P^{t}_i$ can be formulated as shown below,
\begin{align}
\mathrm{(P4):} \quad &  \min_{  P^{t}_i} \;  \frac{D_i\big(\eta   P^{t}_i \!+ \!  P^{h} + P^{r}\big)}{ P^{r} B \log_2 \bigg( 1 + \frac{   P^{t}_i \rho_0}{\sigma^2H^2}\bigg)} \label{Pro4}\\
s.t.
\quad & B \log_2 \Big( 1 + \frac{   P^{t}_i \rho_0}{\sigma^2H^2}\Big) \geq R^{th}  \tag{\ref{Pro4}{a}} \label{Pro4C1}\\
\quad & 0 \leq   P^{t}_i \leq P^{max} \tag{\ref{Pro4}{b}} \label{Pro4C2}
\end{align}
Note that the objective function of (P4) is $T^{h,ini}_i$ and the left hand side (LHS) of \eqref{Pro4C1} is $R^{ini}_i$, which are obtained by introducing Assumption 2 into \eqref{AchRate} and \eqref{VirtualHover}, respectively. Furthermore, problem (P4) is a convex-concave fractional programming with an affine numerator and a concave denominator, as well as convex constraints \cite{shen2018fractional}. Thus, (P4) can be efficiently solved by standard Dinkelbach’s Transform method \cite{dinkelbach1967nonlinear}, which result is denoted by $P^{t,ini^*}_i$. 

\vspace{-0.3cm}
\subsection{Travelling Salesman Problem with energy constraints}
\label{TSPE problem}

Introduce $V_{ij} \! = \! V^{ini^*} $, $\mathbf{h}_i  \!= \!\mathbf{w}_i$  and $P^{t}_i=P^{t,ini^*}_i$ into \eqref{Hoveringenergy}-\eqref{VirtualHover} and  \eqref{flyingenergy2}-\eqref{Virtualfly2}, and denote the results as $E^{h,ini^*}_i$, $T^{h,ini^*}_i$, $E^{f,ini^*}_{ij}$ and $T^{f,ini^*}_{ij}$, all of which are certainly constants once the problem (P3) and (P4) are solved. Accordingly, an initial trajectory can be achieved by solving the following problem,
\begin{align}
\mathrm{(P5):} \; &  \min_{ \mathbf{X}, \mathbf{Z} } \;  \sum_{i \in \mathcal{K} } T^{h,ini^*}_i +  \sum_{(i,j) \in \mathcal{E} } x_{ij} T^{f,ini^*}_{ij} \label{Pro5}\\
s.t.
\quad &  \sum_{j \in \mathcal{K}^a } \! z_{ij} \! - \! \sum_{j \in \mathcal{K}^a } \! z_{ji} = \! \sum_{j \in \mathcal{K}^a} \! \! \big( E^{h,ini^*}_i \! \! + E^{f,ini^*}_{ij} \big) x_{ij} , \forall i \in \mathcal{K}, \tag{\ref{Pro5}{a}} \label{Pro5C1} \\
\quad &  z_{0i} = E^{f,ini^*}_{0i}x_{0i}, \; \forall i \in \mathcal{K}, \tag{\ref{Pro5}{b}} \label{Pro5C2} \\
\quad &  \eqref{TSPE1}-\eqref{TSPE3}, \;  \eqref{TSPE6}-\eqref{TSPE7}, \tag{\ref{Pro5}{c}} \label{Pro5C3}
\end{align}
Problem (P5) is significantly easier to be solved to optimality  than (P2) since the objective function and all constraints in (P5) are linear. Furthermore, we have the following proposition for problem (P5),

\textit{\textbf{Proposition 2:} The TSPE problem (P5) is NP-hard.}

\textit{Proof:} Let the constraints in \eqref{TSPE6} are relaxed, that is, $F^{max} \! \! \to \! \! \infty$. It can be easily verified that the optimal solution of problem (P5) must include exactly one sub-tour (Hamiltonian cycle), since the time consumption in the objective function of (P5) satisfies the triangle inequality. Then, the proposed problem (P5) is analogous to the TSP, which is a well-known NP-hard problem. Therefore, with finite battery capacity $F^{max}$, the TSPE problem (P5) is reducible to the TSP. This completes the proof of Proposition 2. $\square$

Although (P5) is NP-hard, several optimization tools, such as Gurobi \cite{gurobi}, could provide an optimal solution for (P5) efficiently for small to medium size problem instances. Notably, the running time of optimization tools mainly depends on the problem formulation. The simulation results in section \ref{NumericalResults} show a significant gain of our flow-based formulation when compared with another formulation recently proposed in \cite{yao2019qos2}. Moreover, a heuristic method with low complexity, named as clone searching algorithm (CSA), is proposed in \cite{yao2019qos2} to obtain a sub-optimal solution when the problem scale is large. 

\vspace{-0.2cm}
\section{Velocity, Transmit Power and Hovering Position Optimization}
\label{VTPandHPopt}

Based on the initial trajectory design as described in section \ref{initializing}, we will now focus on minimizing the total time consumption for each sub-tour by optimizing the velocity, transmit power and hovering positions in this section. Specifically, we select a certain sub-tour and denote all GTs visited in this sub-tour and RS as $\mathcal{K}^{a,s} \subseteq \mathcal{K}^a$. The edges travelled in this sub-tour are denoted by the set $\mathcal{E}^{s} \triangleq \{(i,j)\big| \, i,j  \in \mathcal{K}^{a,s}, x_{ij} = 1\}$, where $x_{ij}$ is the binary variables solved by problem (P4). For notational convenience, we denote $\mathcal{K}^{s} \triangleq \mathcal{K}^{a,s} - \{0\}$ as all GTs involved by $\mathcal{K}^{a,s}$. According to the original problem (P1), we have the following formulation to optimize velocity, transmit power and hovering positions for the sub-tour,
\begin{align}
\mathrm{(P6):} \; &  \min_{\mathbf{H}^s, \mathbf{P}^s, \mathbf{V}^s}  \sum_{i \in \mathcal{K}^{s} } \! \frac{D_i\big(\eta  P^{t}_i \!+ \!  P^{h}  + P^{r}\big)}{ P^{r} B\log_2 \Big( 1 + \frac{   P^{t}_i \rho_0}{\sigma^2(H^2 + \lVert \mathbf{h}_i - \mathbf{w}_i \rVert^2)}\Big)} \notag \\
& +  \sum_{(i,j) \in \mathcal{E}^{s} } \lVert\mathbf{h}_i-\mathbf{h}_j\rVert \Big( \tau_1 V_{ij}^{2} + \tau_2 V_{ij} + \frac{\tau_3}{V_{ij}} + \frac{\tau_4}{V_{ij}^2}  \Big) \label{Pro6}\\
s.t.
\quad & B\log_2 \Big( 1 + \frac{   P^{t}_i \rho_0}{\sigma^2(H^2 + \lVert \mathbf{h}_i \! - \! \mathbf{w}_i \rVert^2)}\Big) \! \geq R^{th}, \, \forall i \in \mathcal{K}^{s} \tag{\ref{Pro6}{a}} \label{Pro6C1}\\
\quad & 0 \leq   P^{t}_i \leq P^{max},  \;  \forall i \in \mathcal{K}^{s}  \tag{\ref{Pro6}{b}} \label{Pro6C2}\\
\quad & 0 \leq V_{ij} \leq V^{max} , \;  \forall (i,j) \in \mathcal{E}^s \tag{\ref{Pro6}{c}}  \label{Pro6C3}\\
\quad & \sum_{(i,j) \in \mathcal{E}^{s} } \! \lVert\mathbf{h}_i \! - \! \mathbf{h}_j\rVert \Big( \psi_1 V_{ij}^{2} + \psi_2 V_{ij} + \frac{\psi_3}{V_{ij}} + \frac{\psi_4}{V_{ij}^2} \Big) \notag \\
\quad &  + \sum_{i \in \mathcal{K}^{s} } \! \frac{D_i \big( \eta   P^{t}_i  \! +\!  P^{h} \big)}{B\log_2 \Big( 1 + \frac{   P^{t}_i \rho_0}{\sigma^2(H^2 + \lVert \mathbf{h}_i - \mathbf{w}_i \rVert^2)}\Big)}
\leq F^{max} \tag{\ref{Pro6}{d}} \label{Pro6C4} \\
\quad & \mathbf{h}_0 = \mathbf{w}_0 \tag{\ref{Pro6}{e}} \label{Pro6C5} 
\end{align}
where $\mathbf{H}^s \triangleq \{\mathbf{h}_i \big| \, \forall i\in \mathcal{K}^{a,s}\}$, $\mathbf{P}^s \triangleq \{  P^{t}_i \big | \, \forall i \in \mathcal{K}^s \}$ and  $\mathbf{V}^s \triangleq \{V_{ij} \big|\, \forall (i,j) \in \mathcal{E}^s\}$ are the variables in the chosen sub-tour. Compared with the problem (P1), the TSPE energy constraints \eqref{TSPE4}-\eqref{TSPE6} are replaced by \eqref{Pro6C4} since the order of UAV visits is known. The main difficulty for optimally solving (P6) lies in the fractional parts in the objective function and constraints \eqref{Pro6C4}. To tackle this difficulty, by introducing slack variables $\mathbf{S}^s \triangleq \{S_i \big| \, S_i  \! \geq \! 0, i  \in  \mathcal{K}^{s} \}$, problem (P6) can be rewritten as,
\begin{align}
\mathrm{(P6.1):} \; &  \min_{\mathbf{H}^s, \mathbf{P}^s, \mathbf{V}^s,\mathbf{S}^s}  \sum_{i \in \mathcal{K}^{s} } \! \frac{D_i\big(\eta  P^{t}_i \!+ \!  P^{h}  + P^{r}\big)}{ P^{r} S_i} \notag \\
& +  \sum_{(i,j) \in \mathcal{E}^{s} } \lVert\mathbf{h}_i-\mathbf{h}_j\rVert \Big( \tau_1 V_{ij}^{2} + \tau_2 V_{ij} + \frac{\tau_3}{V_{ij}} + \frac{\tau_4}{V_{ij}^2}  \Big) \label{Pro61}\\
s.t.
\quad & B\log_2 \Big( 1 + \frac{   P^{t}_i \rho_0}{\sigma^2(H^2 + \lVert \mathbf{h}_i \! - \! \mathbf{w}_i \rVert^2)}\Big) \! \geq R^{th}, \, \forall i \in \mathcal{K}^{s} \tag{\ref{Pro61}{a}} \label{Pro61C1}\\
\quad & 0 \leq   P^{t}_i \leq P^{max},  \;  \forall i \in \mathcal{K}^{s}  \tag{\ref{Pro61}{b}} \label{Pro61C2}\\
\quad & 0 \leq V_{ij} \leq V^{max} , \;  \forall (i,j) \in \mathcal{E}^s \tag{\ref{Pro61}{c}}  \label{Pro61C3}\\
\quad & \sum_{(i,j) \in \mathcal{E}^{s} } \! \lVert\mathbf{h}_i \! - \! \mathbf{h}_j\rVert \Big( \psi_1 V_{ij}^{2} + \psi_2 V_{ij} + \frac{\psi_3}{V_{ij}} + \frac{\psi_4}{V_{ij}^2} \Big) \notag \\
\quad & \qquad \qquad \quad + \sum_{i \in \mathcal{K}^{s} } \! \frac{D_i \big( \eta   P^{t}_i  \! +\!  P^{h} \big)}{S_i}
\leq F^{max} \tag{\ref{Pro61}{d}} \label{Pro61C4} \\
\quad & \mathbf{h}_0 = \mathbf{w}_0 \tag{\ref{Pro61}{e}} \label{Pro61C5} \\
\quad & B\log_2 \Big( 1 + \frac{   P^{t}_i \rho_0}{\sigma^2(H^2 + \lVert \mathbf{h}_i - \mathbf{w}_i \rVert^2)}\Big) \geq S_i, \; \forall i \in \mathcal{K}^{s} \tag{\ref{Pro61}{f}} \label{Pro61C6} \\
\quad &  S_i \geq 0, \;  \forall i \in \mathcal{K}^{s} \tag{\ref{Pro61}{g}} \label{Pro61C7}
\end{align}
It can be easily veriried that all constraints in \eqref{Pro61C6} can be met with equality when the problem (P6.1) is solved optimally, since otherwise $S_i$ can be increased to reduce the value of objective function without breaking the constraints in \eqref{Pro61C4}. However, problem (P6.1) is still not solvable since the non-convex objective function and constraints \eqref{Pro61C1}, \eqref{Pro61C4} and \eqref{Pro61C6}. In the following, based on the block coordinate descent (BCD) method, we partition the variables of (P6.1) into two groups, $(\mathbf{V}^s, \mathbf{P}^s)$ and $(\mathbf{H}^s, \mathbf{S}^s)$, and solve them iteratively.

\vspace{-0.3cm}
\subsection{Velocity and Transmit Power Optimization}
\label{VelocityOptimization}
For any given $\mathbf{H}^s$ and $\mathbf{S}^s$, $\mathbf{V}^s$ and $\mathbf{P}^s$ in (P6.1) can be optimized jointly by the following problem,
\begin{align}
\mathrm{(P7):} \; &  \min_{\ \mathbf{P}^s, \mathbf{V}^s}  \sum_{i \in \mathcal{K}^{s} } \! \frac{D_i\big(\eta  P^{t}_i \!+ \!  P^{h}  + P^{r}\big)}{ P^{r} S_i}  + \notag \\
&  \sum_{(i,j) \in \mathcal{E}^{s} } \lVert\mathbf{h}_i-\mathbf{h}_j\rVert \Big( \tau_1 V_{ij}^{2} + \tau_2 V_{ij} + \frac{\tau_3}{V_{ij}} + \frac{\tau_4}{V_{ij}^2}  \Big)  \label{Pro7}\\
s.t.
\quad &  B\log_2 \Big( 1 + \frac{   P^{t}_i \rho_0}{\sigma^2(H^2 + \lVert \mathbf{h}_i - \mathbf{w}_i \rVert^2)}\Big) \! \geq R^{th}, \, \forall i \in \mathcal{K}^{s} \tag{\ref{Pro7}{a}} \label{Pro7C1}\\
\quad & 0 \leq   P^{t}_i \leq P^{max},  \;  \forall i \in \mathcal{K}^{s}  \tag{\ref{Pro7}{b}} \label{Pro7C2}\\
\quad & 0 \leq V_{ij} \leq V^{max} , \;  \forall (i,j) \in \mathcal{E}^s \tag{\ref{Pro7}{c}}  \label{Pro7C3}\\
\quad & \sum_{(i,j) \in \mathcal{E}^{s} } \! \lVert\mathbf{h}_i-\mathbf{h}_j\rVert \Big( \psi_1 V_{ij}^{2} + \psi_2 V_{ij} + \frac{\psi_3}{V_{ij}} + \frac{\psi_4}{V_{ij}^2} \Big) \notag \\
\quad & \qquad \qquad \qquad + \sum_{i \in \mathcal{K}^{s} } \frac{D_i \big( \eta   P^{t}_i  \! +\!  P^{h} \big)}{S_i}
\leq F^{max} \tag{\ref{Pro7}{d}} \label{Pro7C4} \\
\quad & B\log_2 \Big( 1 + \frac{   P^{t}_i \rho_0}{\sigma^2(H^2 + \lVert \mathbf{h}_i - \mathbf{w}_i \rVert^2)}\Big) \geq S_i, \; \forall i \in \mathcal{K}^{s} \tag{\ref{Pro7}{e}} \label{Pro7C5}
\end{align}
It can be easy verified that problem (P7) is a convex optimization so that it could be solved efficiently by CVX \cite{cvx}.

\vspace{-0.3cm}
\subsection{Hovering Position and Slack Variable Optimization}
\label{HPOpt}
For any given $\mathbf{V}^s$ and $\mathbf{P}^s$, $\mathbf{H}^s$ and $\mathbf{S}^s$ the underlying optimization problem can be formulated as follows,
\begin{align}
\mathrm{(P8):} \; &  \min_{\mathbf{H}^s,\mathbf{S}^s}  \sum_{i \in \mathcal{K}^{s} } \! \frac{D_i\big(\eta  P^{t}_i \!+ \!  P^{h}  + P^{r}\big)}{ P^{r} S_i}  \notag \\ 
& + \sum_{(i,j) \in \mathcal{E}^{s} } \lVert\mathbf{h}_i-\mathbf{h}_j\rVert \Big( \tau_1 V_{ij}^{2} + \tau_2 V_{ij} + \frac{\tau_3}{V_{ij}} + \frac{\tau_4}{V_{ij}^2}  \Big)  \label{Pro8}\\
s.t.
\quad & B\log_2 \Big( 1 + \frac{   P^{t}_i \rho_0}{\sigma^2(H^2 + \lVert \mathbf{h}_i - \mathbf{w}_i \rVert^2)}\Big) \! \geq R^{th}, \, \forall i \in \mathcal{K}^{s} \tag{\ref{Pro8}{a}} \label{Pro8C1}\\
\quad & \sum_{(i,j) \in \mathcal{E}^{s} } \! \lVert\mathbf{h}_i-\mathbf{h}_j\rVert \Big( \psi_1 V_{ij}^{2} + \psi_2 V_{ij} + \frac{\psi_3}{V_{ij}} + \frac{\psi_4}{V_{ij}^2} \Big) \notag \\
\quad & \qquad \qquad \qquad + \sum_{i \in \mathcal{K}^{s} } \! \frac{D_i \big( \eta   P^{t}_i  \! +\!  P^{h} \big)}{S_i}
\leq F^{max} \tag{\ref{Pro8}{d}} \label{Pro8C2} \\
\quad & \mathbf{h}_0 = \mathbf{w}_0 \tag{\ref{Pro8}{c}} \label{Pro8C3} \\
\quad & B\log_2 \Big( 1 + \frac{   P^{t}_i \rho_0}{\sigma^2(H^2 + \lVert \mathbf{h}_i - \mathbf{w}_i \rVert^2)}\Big) \geq S_i, \; \forall i \in \mathcal{K}^{s} \tag{\ref{Pro8}{d}} \label{Pro8C4} \\
\quad &  S_i \geq 0, \;  \forall i \in \mathcal{K}^{s} \tag{\ref{Pro8}{e}} \label{Pro8C5}
\end{align}
Problem (P8) is still difficult to attack due to the non-convex constraints \eqref{Pro8C1} and \eqref{Pro8C4}. However, observe that although the LHS of \eqref{Pro8C1} and \eqref{Pro8C4} is not concave with respect to $\mathbf{h}_i$, it is convex with respect to $\lVert \mathbf{h}_i - \mathbf{w}_i \rVert^2$. It is well-known that any convex function is globally lower-bounded by its first-order Taylor expansion at any point. Thus, a lower bound of the LHS of \eqref{Pro8C1} and \eqref{Pro8C4} can be obtained as,
\begin{equation}
\label{LowerBound}
\begin{aligned}
& \quad B\log_2 \Big( 1 + \frac{P^{t}_i \rho_0}{\sigma^2(H^2 + \lVert \mathbf{h}_i - \mathbf{w}_i \rVert^2)}\Big) \\
& \geq -C_{1,i}[k] \Big( \lVert \mathbf{h}_i - \mathbf{w}_i \rVert^2 -\lVert \mathbf{h}_i[k] - \mathbf{w}_i \rVert^2 \Big) + C_{2,i}[k] \\
& \triangleq R^{lb}_i
\end{aligned} 
\end{equation}
where $\mathbf{h}_i[k]$ is the local point obtained in $k^{th}$ iteration, and $C_{1,i}[k]$ and $C_{2,i}[k]$ are constants given by,
\begin{equation}
\label{LBparameter1}
\begin{aligned}
C_{1,i}[k] = \frac{\frac{B\sigma^2P^{t}_i \rho_0}{\big(\sigma^2(H^2 + \lVert \mathbf{h}_i[k] - \mathbf{w}_i \rVert^2)\big)^2} \log_2(e)}{1 + \frac{P^{t}_i \rho_0}{\sigma^2(H^2 + \lVert \mathbf{h}_i[k] - \mathbf{w}_i \rVert^2)}} \geq 0,
\end{aligned} 
\end{equation}
\begin{equation}
\label{LBparameter2}
\begin{aligned}
C_{2,i}[k] = B\log_2 \Big( 1 + \frac{P^{t}_i \rho_0}{\sigma^2(H^2 + \lVert \mathbf{h}_i[k] - \mathbf{w}_i \rVert^2)}\Big),
\end{aligned} 
\end{equation}
Since the coefficient $C_{1,i}[k] \geq 0$, $R^{lb}_i$ is certainly concave with respect to $\mathbf{h}_i$. Thus, problem (P8) can be approximated as a convex form, that is, \begin{align}
\mathrm{(P8.1):} \; &  \min_{\mathbf{H}^s,\mathbf{S}^s}  \sum_{i \in \mathcal{K}^{s} } \! \frac{D_i\big(\eta  P^{t}_i \!+ \!  P^{h}  + P^{r}\big)}{ P^{r} S_i}  \notag \\ 
& + \! \sum_{(i,j) \in \mathcal{E}^{s} } \lVert\mathbf{h}_i-\mathbf{h}_j\rVert \Big( \tau_1 V_{ij}^{2} + \tau_2 V_{ij} + \frac{\tau_3}{V_{ij}} + \frac{\tau_4}{V_{ij}^2}  \Big)  \label{Pro81}\\
s.t.
\quad & R^{lb}_i \geq R^{th}, \, \forall i \in \mathcal{K}^{s} \tag{\ref{Pro81}{a}} \label{Pro81C1}\\
\quad & \sum_{(i,j) \in \mathcal{E}^{s} } \! \lVert\mathbf{h}_i-\mathbf{h}_j\rVert \Big( \psi_1 V_{ij}^{2} + \psi_2 V_{ij} + \frac{\psi_3}{V_{ij}} + \frac{\psi_4}{V_{ij}^2} \Big) \notag \\
\quad & \qquad \qquad \quad + \sum_{i \in \mathcal{K}^{s} } \! \frac{D_i \big( \eta P^{t}_i  \! +\!  P^{h} \big)}{S_i}
\leq F^{max} \tag{\ref{Pro81}{b}} \label{Pro81C2} \\
\quad & \mathbf{h}_0 = \mathbf{w}_0 \tag{\ref{Pro81}{c}} \label{Pro81C3} \\
\quad & R^{lb}_i \geq S_i, \; \forall i \in \mathcal{K}^{s} \tag{\ref{Pro81}{d}} \label{Pro81C4} \\
\quad &  S_i \geq 0, \;  \forall i \in \mathcal{K}^{s} \tag{\ref{Pro81}{e}} \label{Pro81C5}
\end{align}
Recalling the successive convex approximation (SCA) method \cite{razaviyayn2014successive}, solving the convex problem (P8.1) iteratively can converge to a locally optimal solution to problem (P8) that satisfies the Karush-Kuhn-Tucker (KKT) conditions.

\vspace{-0.3cm}
\subsection{Overall Algorithm and Convergence}
\label{Convergence}

\begin{algorithm}[!t]
\caption{Velocity, transmit power and hovering position optimization }
\label{optalg}
\begin{algorithmic}[1]
\STATE Obtain the initialized trajectory design by solving problems (P3)-(P5).
\FORALL{sub-tour in the initial trajectory}
\STATE Adopting the results of (P3)-(P4) to initialize the variables $\mathbf{H}^s[0]$, $\mathbf{P}^s[0]$ and $\mathbf{V}^s[0]$. Initialize the slack variables as $\mathbf{S}^{s}[0] \! = \! \{S_i[0] \big| \, S_i[0] = B\log_2 \big( 1 + \frac{ P^{t}_i[0] \rho_0}{\sigma^2H^2}\big), \, i \in  \mathcal{K}^{s}\}$. Let $k=0$.
\REPEAT
\STATE Solve problem (P7) with given $\mathbf{H}^s[k]$ and $\mathbf{S}^s[k]$, denote the solution as $\mathbf{P}^s[k+1]$ and $\mathbf{V}^s[k+1]$.
\STATE Solve problem (P8.1) with given $\mathbf{P}^s[k+1]$ and $\mathbf{V}^s[k+1]$, denote the solution as $\mathbf{H}^s[k+1]$ and $\mathbf{S}^s[k+1]$.
\STATE Update k = k+1.
\UNTIL{The fractional decrease of the objective value is below a predefined threshold.}
\ENDFOR
\end{algorithmic}
\end{algorithm}

According to the results presented in section \ref{VelocityOptimization} and section \ref{HPOpt}, an overall iterative algorithm for problem (P6.1) is proposed as Algorithm \ref{optalg}. It is worth pointing out that although the convergence of BCD method is guaranteed, Algorithm \ref{optalg} differs from classic BCD process since an approximation sub-problem (P8.1) is solved in step 6 instead of the exact sub-problem (P8). The convergence analysis is shown as the following proposition and simulation results of the convergence behaviour can be found in section \ref{NumericalResults}.

\textit{\textbf{Proposition 3:} Algorithm \ref{optalg} is guaranteed to converge. }

\textit{Proof:} According to the previous convergence analysis in \cite{wu2018joint}, the objective value of (P6.1) is proven to be non-increasing after each iteration of Algorithm \ref{optalg}, that is,
\begin{equation}
\label{convergence}
\begin{aligned}
& \; \Phi^{P6.1}(\mathbf{H}^{s}[k+1], \mathbf{P}^{s}[k+1], \mathbf{V}^{s}[k+1],\mathbf{S}^{s}[k+1]) \\
\leq & \; \Phi^{P6.1}(\mathbf{H}^{s}[k], \mathbf{P}^{s}[k], \mathbf{V}^{s}[k],\mathbf{S}^{s}[k])
\end{aligned} 
\end{equation} 
Besides, the objective value $\Phi^{P6.1}(\mathbf{H}^s, \mathbf{P}^s, \mathbf{V}^s,\mathbf{S}^s)$ is lower-bounded at least by 0. Thus, Algorithm \ref{optalg} must be converged to a certain point with a finite number of iterations. This is completes the proof of Proposition 3. $\square$

\vspace{-0.2cm}
\section{Numerical Results}
\label{NumericalResults}

\begin{table}[!t]
\centering
\caption{Parameter Settings}
\label{TAB para}
\begin{tabular}{ll|ll}
\hline
Parameter & Value & Parameter & Value\\
\hline
$B$ & 2 MHz & $\sigma^2$ & -110 dBm\\
$H$ & 100 m & $\rho_0$ & -60 dB \\
$R_{th}$  & 15 Mbps & $P^r$ &100 W\\
$F^{max}$ & 399600 J \cite{jaafar2019dynamics} & $V^{max}$ & 25 m/s \\
$K$  & 20 &  $D_i$ & 1 Gbits (except Fig.\ref{performances}) \\
$\eta$ & 4.2 \cite{yao2019qos} & $P^h$ & 165 W \cite{zeng2019energy}\\ 
$[\tau_1,\tau_2]$ & $[0.9,1.6]\! \times \! 10^{-4}$ \cite{zeng2019energy} & $[\psi_1,\psi_2]$ & $[9.2,16.6] \! \times \! 10^{-3}$ \cite{zeng2019energy}  \\
$[\tau_3,\tau_4]$ & $[1.8,3.6]$ \cite{zeng2019energy} & $[\psi_3,\psi_4]$ &$[79.9,357.2]$ \cite{zeng2019energy} \\
\hline
\end{tabular}
\vspace{-0.3cm}
\end{table}

\begin{figure*}[!t]
\centering 
\subfigure['EFFI' trajectory design, total time = $9931.54$ s]{ \label{effi_trajectory} 
\includegraphics[width=.42\textwidth]{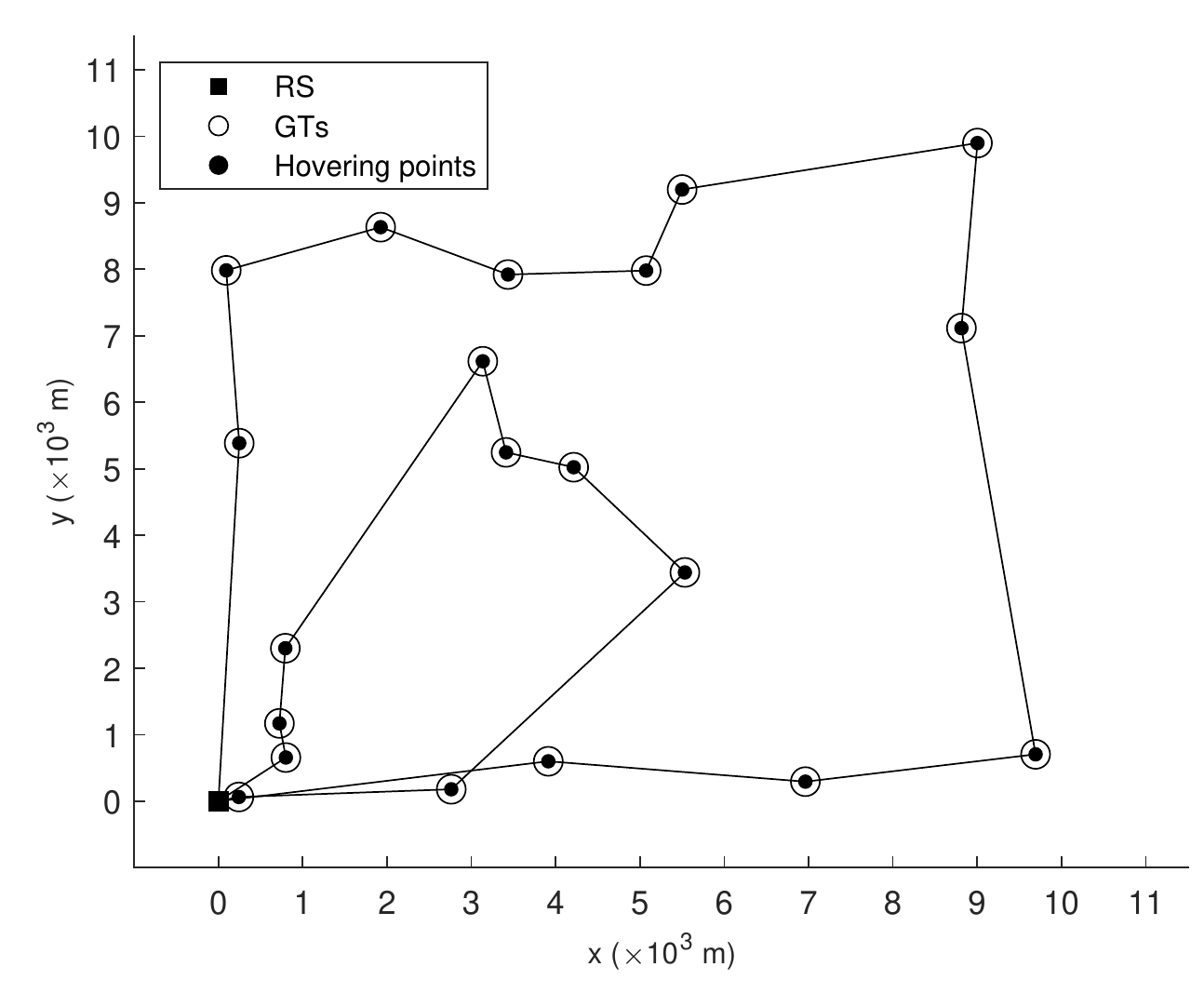}} 
\subfigure['MAX' trajectory design, total time = $9874.73$ s]{ 
\label{max_trajectory} 
\includegraphics[width=.42\textwidth]{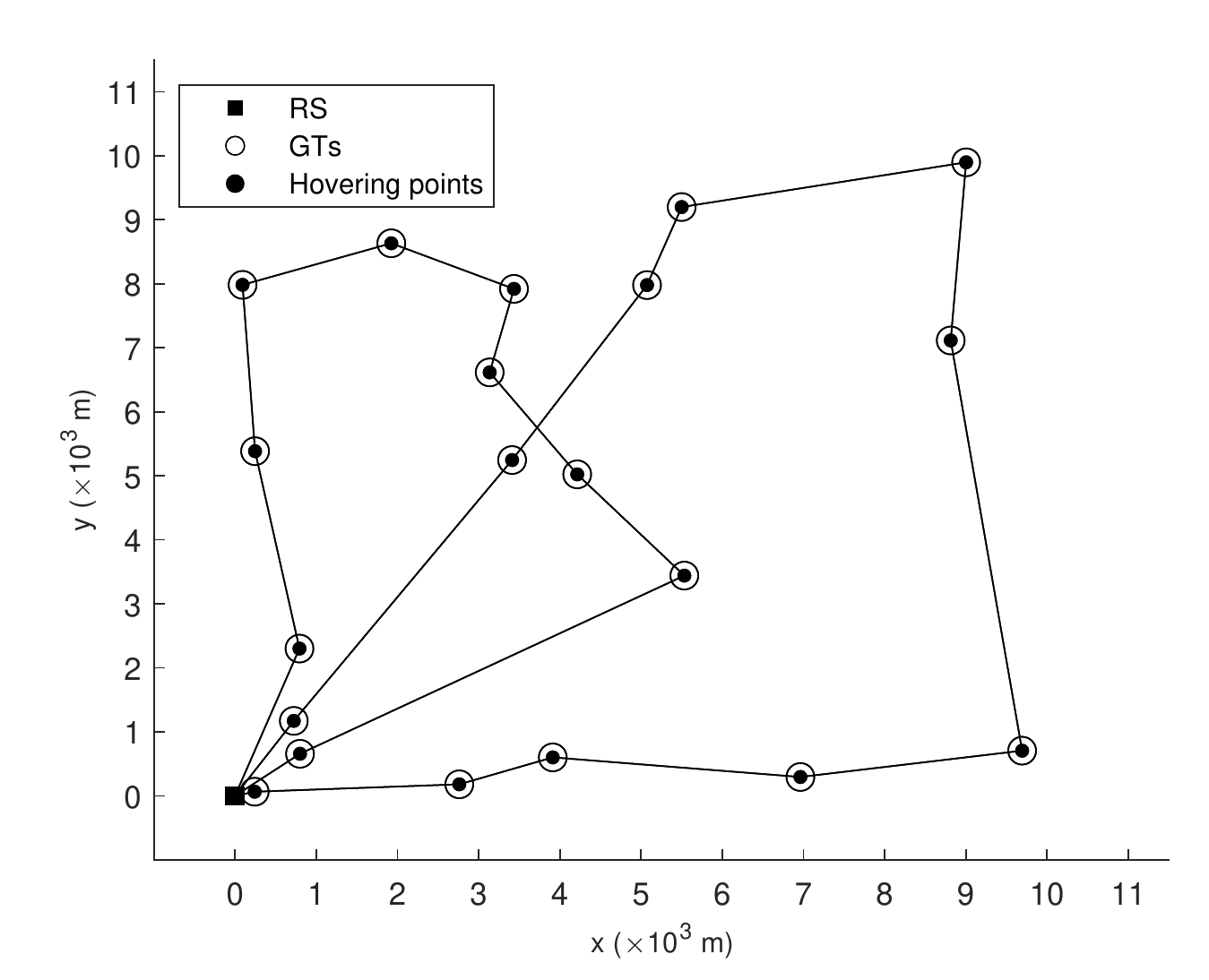}} 
\quad
\subfigure['INI' trajectory design, total time = $9678.02$ s]{ 
\label{ini_trajectory} 
\includegraphics[width=.42\textwidth]{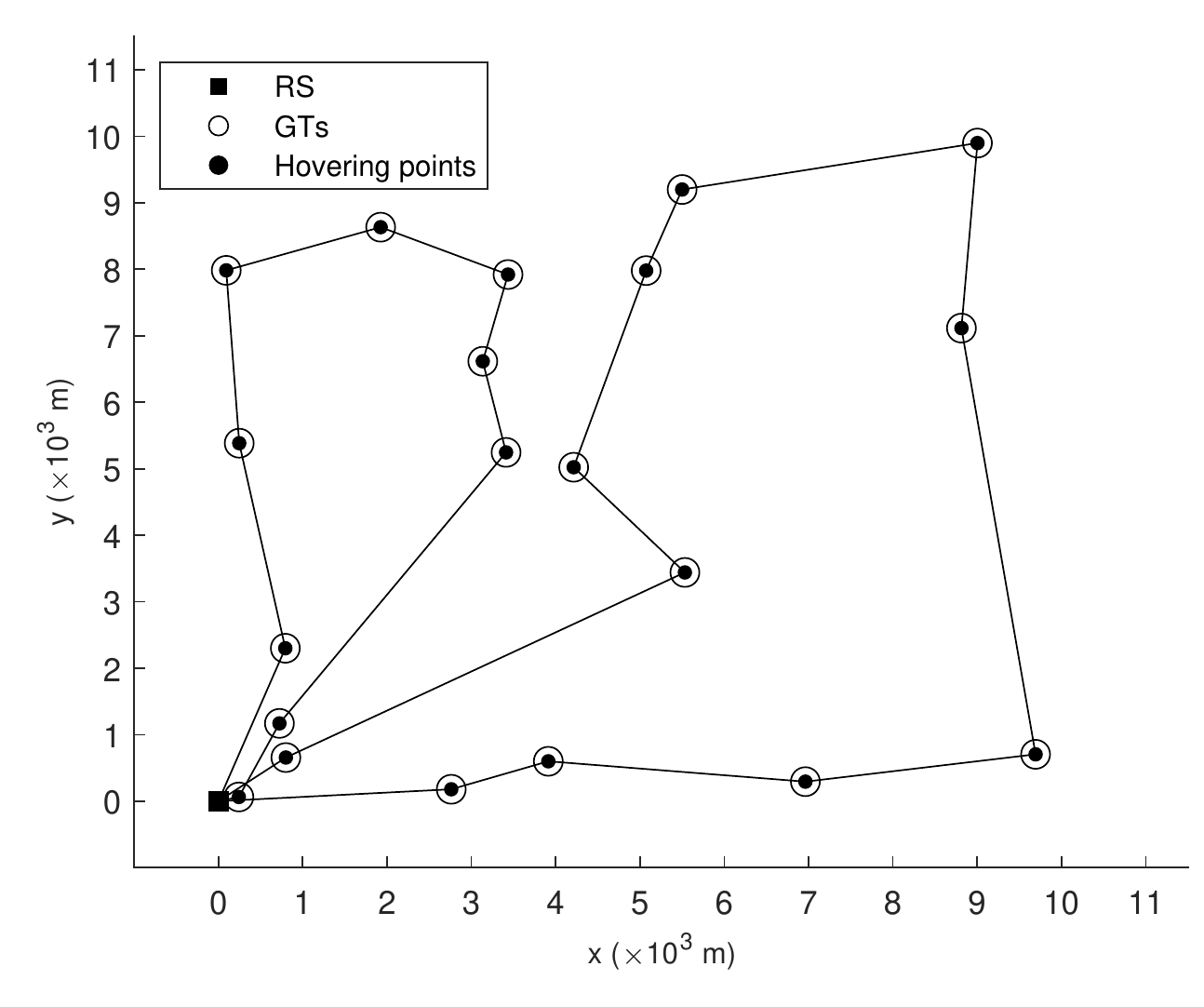}} 
\subfigure['OPT' trajectory design, total time = $9495.68$ s]{ 
\label{opt_trajectory} 
\includegraphics[width=.42\textwidth]{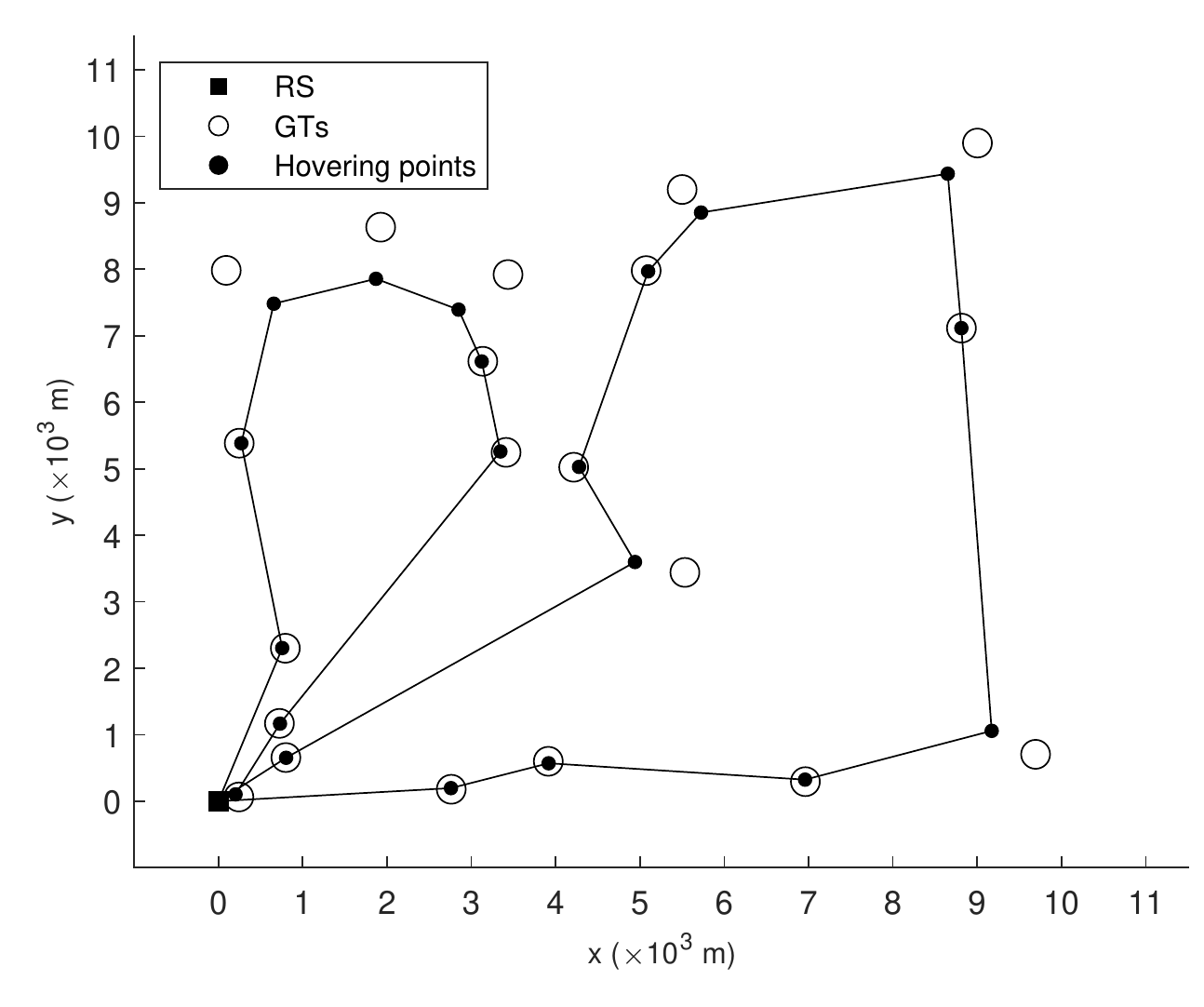}} 
\caption{Comparison of the trajectories with different designs and their total time consumption}
\label{ComTrajectory} 
\vspace{-0.3cm}
\end{figure*}

In this section, numerical results are presented to evaluate the performance of the proposed method. The parameter settings are summarized in Table \ref{TAB para}. Note that the battery capacity is generally measured in Ampere-seconds ($\mathrm{A  s}$) under a certain output voltage. For computing convenience, we calculate the capacity of a 3-cell (3S) LiPo battery by $11.1 \, \mathrm{V} \times 36000 \, \mathrm{As} = 399600 \, \mathrm{J}$  in this section \cite{jaafar2019dynamics}. We mainly compare the following four trajectory designs,
\begin{itemize}
    \item{'EFFI': The UAV flies with the energy-efficient velocity obtained by $V^{ef\!fi} =  \mathop{\arg\min}\limits_{0\leq V \! \leq \! V^{max}} \{ \psi_1 V^{2} + \psi_2 V + \frac{\psi_3}{V} + \frac{\psi_4}{V^2} \}$, where the objective function is the unit energy consumed by $1 \mathrm{m}$ flight. Besides, the UAV would hover right above GTs, that is, $ \{ \mathbf{h}_i \big| \, \mathbf{h}_i \!= \!\mathbf{w}_i, \, \forall i \in \mathcal{K} \}$.}
    \item{'MAX': The UAV always flies with the maximum velocity $V^{max}$ and hovers right above GTs.}
    \item{'INI': The trajectory initialized by problems (P3)-(P5).}
    \item{'OPT': The trajectory optimized by Algorithm \ref{optalg}.}
\end{itemize}

To evaluate the performance of the different schemes in a clear manner, we define the flight time as the summation of both flying and hovering time, which is distinguished from the recharging duration. We then demonstrate three performance metrics: the energy consumption, the flight time and the total time, for different trajectory designs. Notably, as reviewed in section \ref{introduction}, most existing works mostly focus on one of the two above indicators and predominately put emphasis on energy consumption and flight time. However, in the proposed scheme the total time consumption, including the flight and recharging time, is seen as a combined indicator and results hereafter aim to emphasize the gains of the proposed approach. 

Fig.\ref{ComTrajectory} compares four different trajectory designs as well as their total time consumption. Comparing Fig.\ref{effi_trajectory}, Fig.\ref{max_trajectory} and Fig.\ref{ini_trajectory}, it can be observed that the energy efficiency and flight time minimization are coupled with each other. In other words, optimizing any of them alone cannot achieve the optimal total time consumption. Furthermore, comparing Fig.\ref{ini_trajectory} and Fig.\ref{opt_trajectory}, it can be seen that 'OPT' design gains an almost 2\% decrease in time consumption compared with 'INI' strategy.

\begin{figure}[!t]
\centering 
\subfigure[Energy consumption versus data volume]{ 
\label{energyconsumption_sim} 
\includegraphics[width=.425\textwidth]{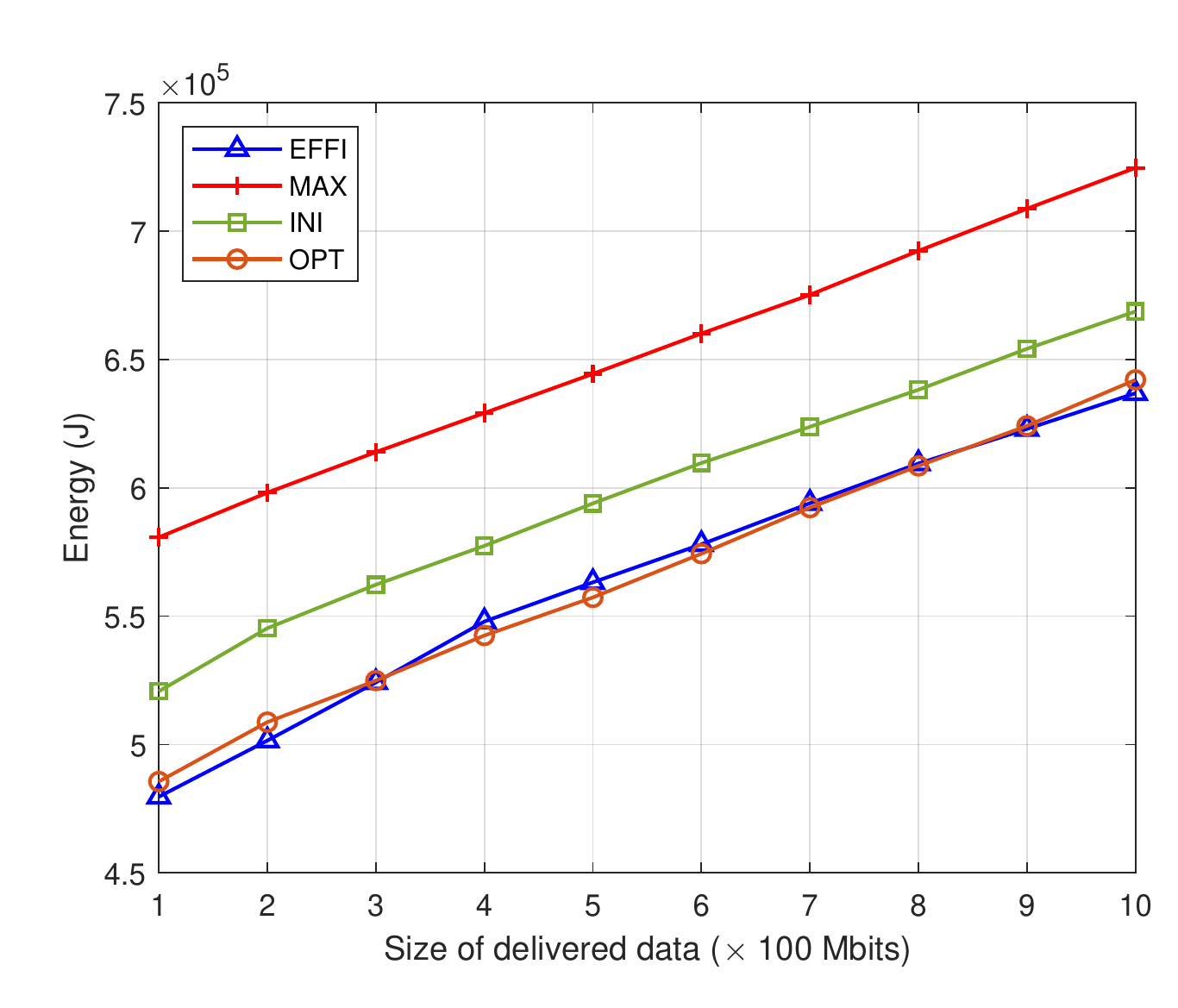}} 
\subfigure[Flight time versus data volume]{ 
\label{flighttime_sim} 
\includegraphics[width=.425\textwidth]{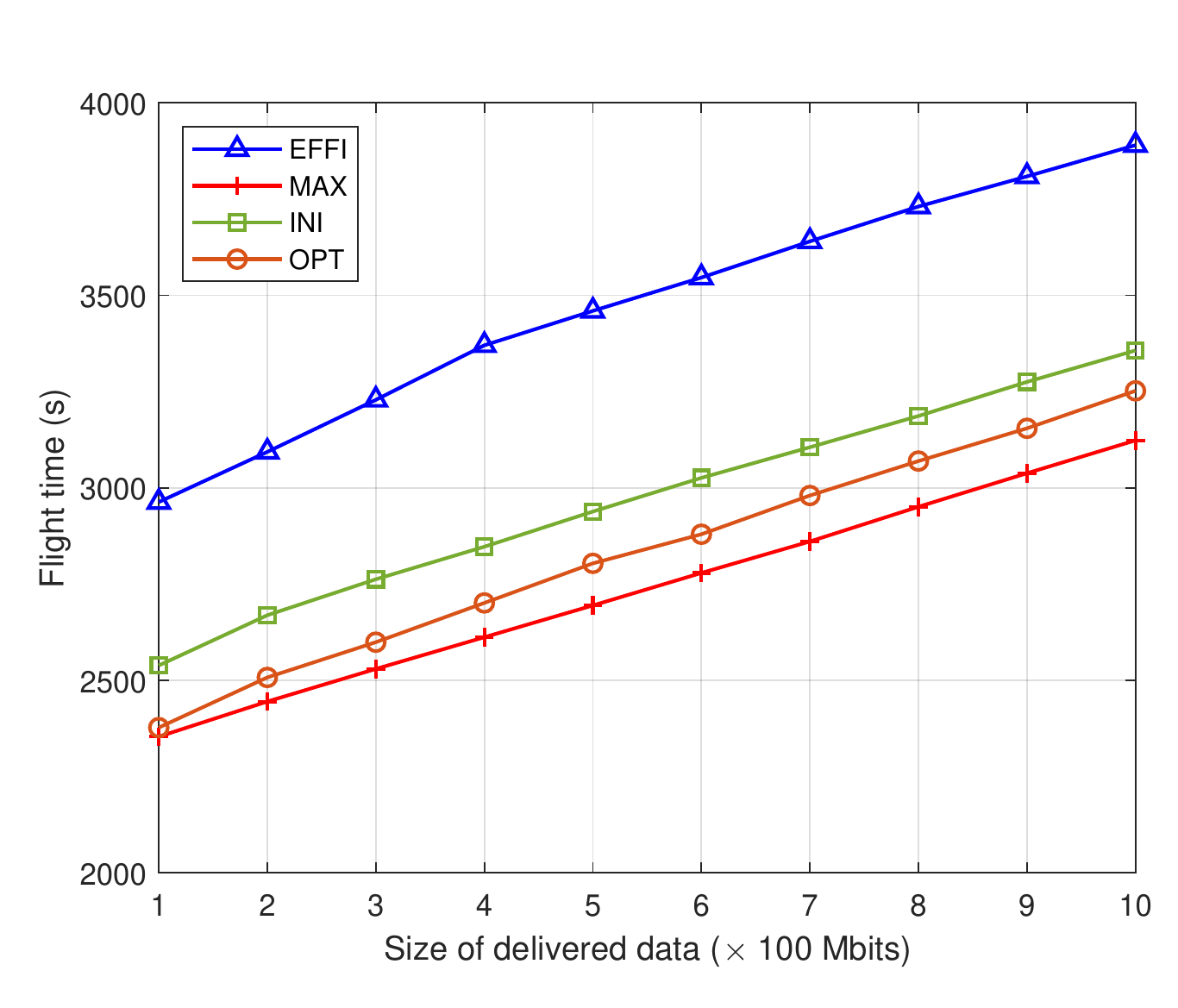}} 
\subfigure[Total time versus data volume]{ 
\label{totaltime_sim} 
\includegraphics[width=.425\textwidth]{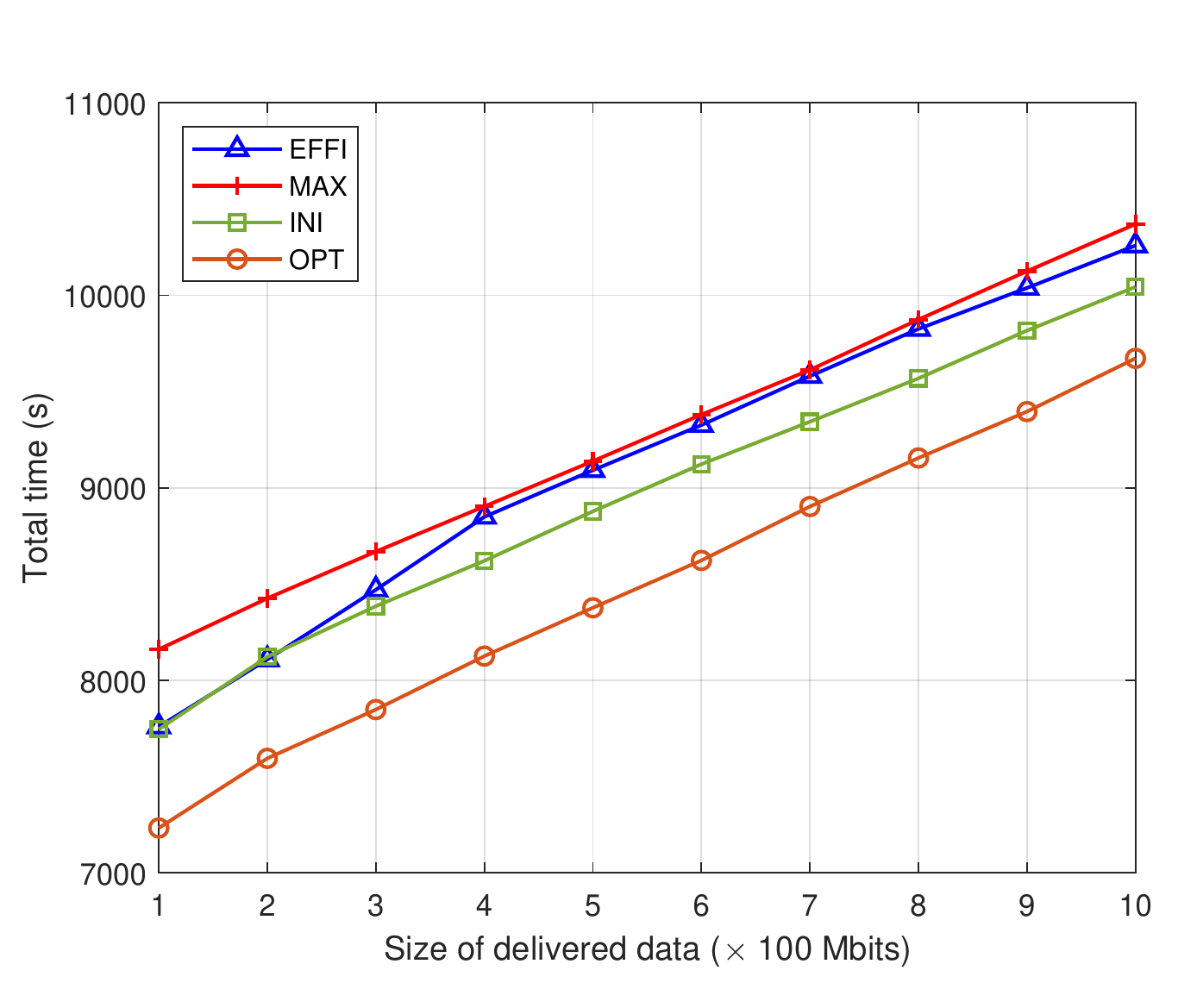}} 
\caption{Performances for different trajectory designs}
\label{performances} 
\vspace{-0.4cm}
\end{figure}

Fig.\ref{performances} compares the four different trajectory designs for different required data volumes under three performance metrics, i.e. energy consumption, flight time and total time, and the results are averaged via 100  Monte Carlo simulations. In Fig.\ref{energyconsumption_sim}, as expected, 'MAX' strategy has the largest energy consumption since there isn't any energy-awareness in the 'MAX' design. However, although the UAV always flies with the most energy-efficient velocity in 'EFFI' trajectory, the 'OPT' strategy shows a nearly equivalent energy efficiency as the 'EFFI' scheme since the UAV does not need to exactly visit the GT locations at the horizontal plane, that is, a shorter flying distance. In Fig.\ref{flighttime_sim}, as expected, the 'MAX' and 'EFFI' strategies show the best and the worst performances in flight time, respectively. In Fig.\ref{totaltime_sim}, it can be observed  that both the 'INI' and 'OPT' trajectories require less total time than benchmarks. For instance, when $D_i = 1$ Gbits, the proposed 'INI' scheme provides nearly 2\% and 3\% gain compared to 'EFFI' and 'MAX', respectively. Also the 'OPT' scheme has an optimality gap of 6\% and 7\% than these two benchmarks.

\begin{figure}[!t]
\centering
\subfigure{\includegraphics[width=.423\textwidth]{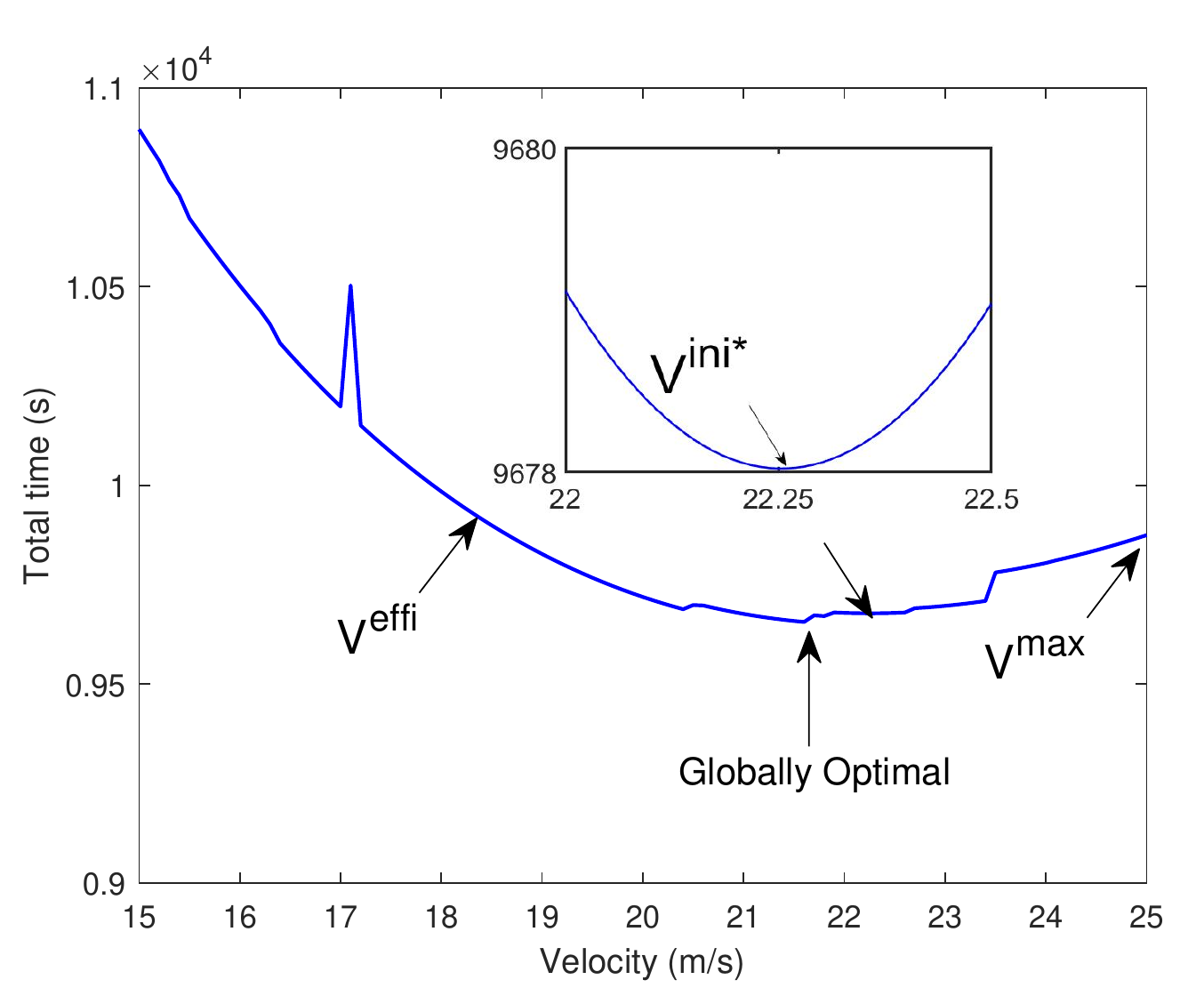}}
\caption{Total time versus different velocity}
\label{differentvelocity}
\vspace{-0.3cm}
\end{figure}

As eluded in section \ref{initializing}, to tackle the difficult problem (P2), we decouple the variables and solve them separately to obtain a near-optimal solution. The following simulation result in Fig.\ref{differentvelocity} shows that the proposed algorithm can achieve a high-quality solution. As expected, when the UAV moves at low velocities it might be the case that not all GT will be visited due to the on-board energy constraint. Thus, we only simulate velocities from 15 m/s to 25 m/s. Besides, the main results in Fig.\ref{differentvelocity} are sampled every 0.1 m/s, whilst results shown in the enclosed sub-figure are sampled every 0.01 m/s. In Fig.\ref{differentvelocity}, it can be observed that the total time is a non-smooth function with respect to the velocity. The explanation is as follows. In the smooth part, the order of UAV visits does not change. Thus, the total time is smooth for changing velocity. However, the order of UAV visits might change because of the on-board battery limitation, which is shown as the sharp increasing or decreasing in Fig.\ref{differentvelocity}. Furthermore, it can also be observed that although the $V^{ini^*}$ obtained by solving (P3) cannot achieve the globally optimal result, it is actually a stationary point, in other words, provides a competitive local optimal solution.

\begin{figure}[!t]
\centering
\subfigure{\includegraphics[width=.42\textwidth]{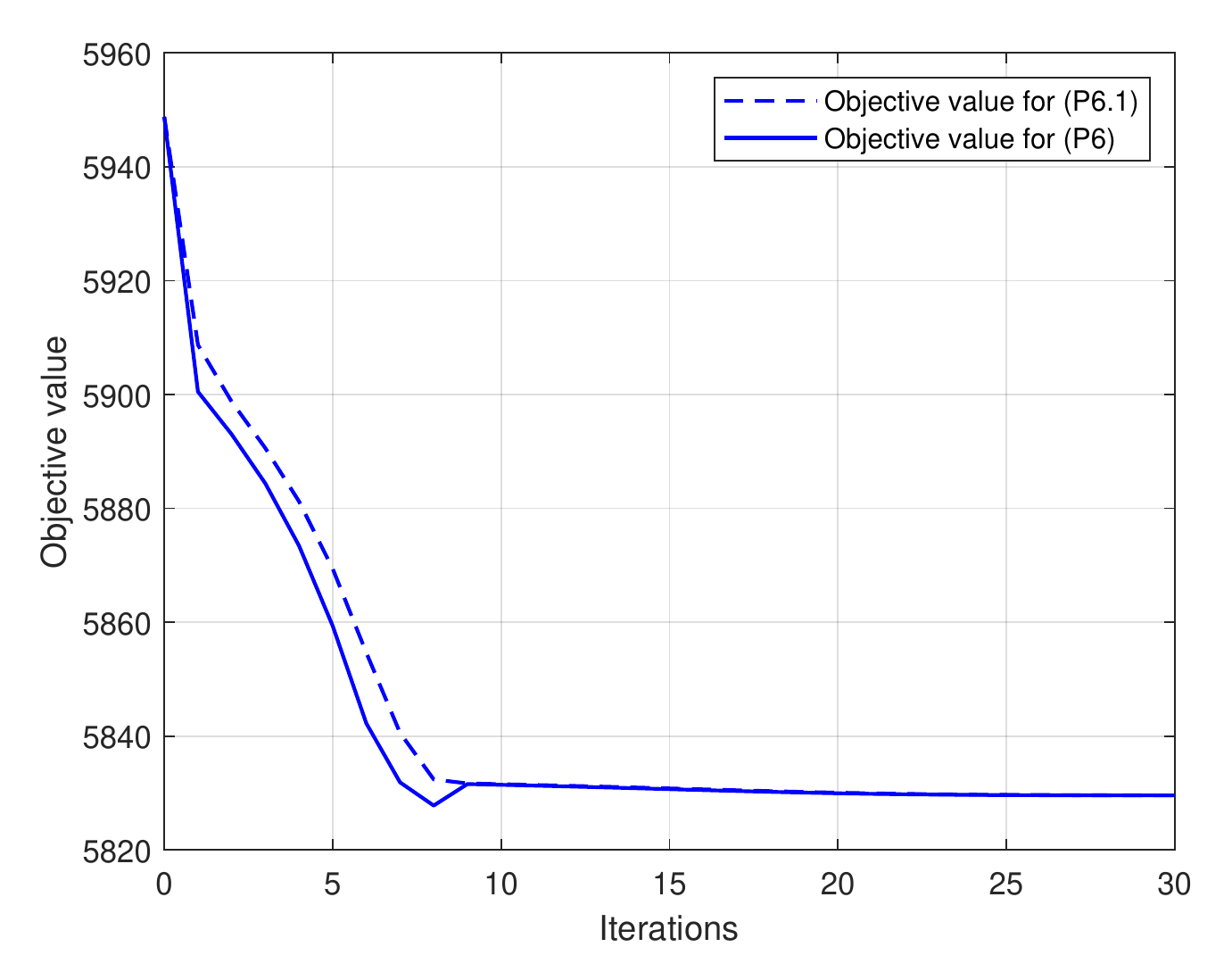}}
\caption{Convergence behaviour for Algorithm \ref{optalg}}
\label{sim_convergence}
\vspace{-0.3cm}
\end{figure}

Fig.\ref{sim_convergence} shows the convergence behaviour when a certain sub-tour is optimized by Algorithm \ref{optalg} in 30 iterations. The solid and dashed lines represent the value of the objective function of problem (P6) and (P6.1), respectively. Since a non-negative slack variable $\mathbf{S}^{s}$ is introduced to problem (P6), the objective value of (P6.1) is in essence an upper bound for problem (P6). This fact is verified and depicted in Fig.\ref{sim_convergence}. Moreover, the analysis in section \ref{Convergence} illustrates the non-increasing of the objective value of problem (P6.1), which is shown by the dashed line in Fig.\ref{sim_convergence}. However, it can be observed that the objective value of (P6) is not always non-increasing, but is eventually converging quickly with finite number of iterations.

The previous work in \cite{yao2019qos2} provides another formulation for TSPE problem, which plays the same role as the flow-based constraint set proposed in section \ref{TSPEM}. It is worth pointing out that although the TSPE problem is NP-hard as proved in Proposition 2, the computation time depends heavily on how the problem is formulated when applying classic algorithms to find the globally optimal solution, such as a branch and cut framework. To evaluate the computation performances, Table \ref{runningtime} compares the running time for solving the same problem with different formulations. The results presented in Table \ref{runningtime} are implemented on MATLAB R2020b and solved by Gurobi 9.9.1\cite{gurobi}, running on a Windows 7 with Intel-i7 2.50 GHz and 8 GB RAM. Table \ref{runningtime} shows that the proposed flow-based formulation has a significant gain compared to previous proposed scheme \cite{yao2019qos2}.

\begin{table}[!t]
\centering
\caption{Running time for different formulations}
\label{runningtime}
\begin{tabular}{m{2cm}<{\centering}|cc}
\hline
Number of GTs K = & Flow-based formulation & Formulation in \cite{yao2019qos2} \\
\hline
10 & 1.53 s & 24239.65 s \\
15 & 6.51 s & 65257.77 s \\
20 & 12.56 s & - \\
30 & 714.20 s & - \\
\hline
\end{tabular}
\vspace{-0.3cm}
\end{table}

\vspace{-0.25cm}
\section{Conclusion}
\label{conclusion}

This paper studies the trajectory design problem for a rechargeable UAV aided wireless network, in which a UAV is deployed to disseminate information to a group of GTs. We unify two important performances, namely, the energy efficiency and the overall flight time, by defining the total time consumption as the sum of recharging and flight time. The total time is then minimized by trajectory planning. Unlike previous research work in the area, we firstly investigate a flow-based formulation  to model  the recharging process and energy capacity of the UAV. We then proposed a two-step method to solve this problem. In the first step, we initialize a feasible trajectory design by fixing the UAV hovering points at the GT horizontal locations. In the second step, based on the initialized trajectory design, the velocity, transmit power and hovering points are optimized jointly to decrease the total time consumption. Numerical results show significant performance gains of the proposed designs over two benchmark schemes, i.e. optimizing the energy consumption or the overall flight time. Moreover, the proposed flow-based formulation also shows a significant gain when comparing the running time with previous proposed type of formulations.

\ifCLASSOPTIONcaptionsoff
  \newpage
\fi

\vspace{-0.25cm}
\bibliographystyle{IEEEtran}
\bibliography{IEEEabrv,reference} 

\end{document}